\documentclass[manuscript]{aastex}

\shorttitle{HF and H$_2$O Survey with {it Herschel}/HIFI}
\shortauthors{Sonnentrucker et al.}

\newcommand{\beq}{\begin{equation}}
\newcommand{\eeq}{\end{equation}}

\begin{document}


\title{A {\it Herschel}/HIFI Legacy Survey of HF and H$_2$O in the Galaxy: \\
    Probing Diffuse Molecular Cloud Chemistry\thanks{{\it Herschel} is an ESA space observatory with science instruments provided by European-led Principal Investigator consortia and with important participation from NASA.}}


\author{P. Sonnentrucker\altaffilmark{1}} 
\affil{Space Telescope Science Institute, Baltimore, MD 21218/European Space Agency}
\email{sonnentr@stsci.edu}

\author{M. Wolfire\altaffilmark{2}}
\affil{University of Maryland, College Park, MD 20742}

\author{D. A. Neufeld\altaffilmark{3}}
\affil{Johns Hopkins University, Baltimore, MD 21218}

\author{N. Flagey\altaffilmark{4}}
\affil{Institute for Astronomy, Hilo, HI 96720-2700 }

\author{M. Gerin\altaffilmark{5,6}}
\affil{Sorbonne Universit\'{e}s, Universit\'{e} Pierre et Marie Curie, Paris 6, CNRS, Observatoire de Paris, UMR 8112, LERMA, Paris, France}
\affil{LERMA, Observatoire de Paris, PSL Research University, CNRS, UMR 8112, F-75014,
Paris, France}

\author{P. Goldsmith\altaffilmark{7}}
\affil{Jet Propulsion Laboratory, 4800 Oak Grove Drive, Pasadena CA 91109-8099}

\author{D. Lis\altaffilmark{5,6,8}}
\affil{Sorbonne Universit\'{e}s, Universit\'{e} Pierre et Marie Curie, Paris 6, CNRS, Observatoire de Paris, UMR 8112, LERMA, Paris, France}
\affil{LERMA, Observatoire de Paris, PSL Research University, CNRS, UMR 8112, F-75014,
Paris, France}

\and 
\author{R. Monje\altaffilmark{8}}
\affil{California Institute of Technology, Cahill Center for Astronomy and Astrophysics 301-17, CA 91125, Pasadena, USA}





\begin{abstract}
We combine {\it Herschel} observations of a total of 12 sources to construct the most uniform survey of HF and H$_2$O in our Galactic disk. Both molecules are detected in absorption along all sight lines. The high spectral resolution of the Heterodyne Instrument for the Far-Infrared (HIFI) allows us to compare the HF and H$_2$O distributions in 47 diffuse cloud components sampling the disk. We find that the HF and H$_2$O velocity distributions follow each other almost perfectly and establish that HF and H$_2$O probe the same gas-phase volume.  Our observations corroborate theoretical predictions that HF is a sensitive tracer of H$_2$ in diffuse clouds, down to molecular fractions of only a few percent. Using HF to trace H$_2$ in our sample, we find that the $N$(H$_2$O)-to-$N$(HF) ratio shows a narrow distribution with a median value of 1.51. Our results further suggest that H$_2$O might be used as a tracer of H$_2$ -within a factor 2.5-  in the diffuse interstellar medium. We show that the measured factor of $\sim$2.5 variation around the median is driven by true local variations in the H$_2$O abundance relative to H$_2$ throughout the disk.  The latter variability allows us to test our theoretical understanding of the chemistry of oxygen-bearing molecules in the diffuse gas. We show that both gas-phase {\it and} grain-surface chemistry are required to reproduce our H$_2$O observations. This survey thus confirms that grain surface reactions can play a significant role in the chemistry occurring in the diffuse interstellar medium ($n_{\rm H}$ $\leq$1000 cm$^{-3}$).
\end{abstract}





\section{Introduction}

Diffuse molecular clouds are regions in which atomic hydrogen is progressively converted to molecular hydrogen (H$_2$), and neutral carbon and carbon monoxide become the dominant forms of carbon as the total visual extinction increases from 0.1 (the onset of H$_2$ formation) to about 2 magnitudes (see, Snow \& McCall 2006, for a review). Diffuse molecular clouds are considered clouds in transition from diffuse mainly atomic gas to fully molecular gas, hence an important first step in star formation. As a result, they play a crucial role in the lifecycle of the interstellar medium making their study critical to advance our understanding of how molecular clouds form from the diffuse interstellar medium (ISM). 

Because of their relatively low densities ($n_{\rm H}$ $\leq$1000 cm$^{-3}$) and their low shielding from UV radiation compared to dense clouds, diffuse molecular clouds were expected to be mostly devoid of molecules. However, the last four decades of UV/optical/radio observations, from space and from the ground, have demonstrated that diffuse molecular clouds have a surprisingly rich and still largely unexplained chemistry (e.g., Snow \& McCall 2006, Sonnentrucker et al. 2007; Sheffer et al. 2008; Liszt et al. 2007; Neufeld et al. 2012).  Comparisons of UV/optical molecular absorption line studies with high-resolution studies using sub-mm data have provided complementary information on the  physics and chemistry of the absorbing gas, over a large range of opacities and at great distances in the Galactic disk. As a result, these clouds constitute ``in-situ" laboratories in which we can study a variety of physical and chemical processes of broad applicability in astrophysics.

Small (diatomic and triatomic) hydrides are important tracers of the diffuse interstellar medium physics and chemistry. Possessing small moments of inertia, these hydrides have rotational transitions at THz frequencies that are difficult or impossible to observe using ground-based observatories. The {\it Herschel}/HIFI Key Program PRISMAS (Probing InterStellar Molecules with Absorption line Studies: PI, M. Gerin) was aimed at surveying key hydrides within the Galaxy. Up to 22 small molecular species were specifically targeted with PRISMAS in order to probe the chemistry of carbon (e.g., Mookerjea et al. 2010, Gerin et al. 2010, Godard et al. 2012; 2014), nitrogen (e.g., Persson et al. 2010; Persson et al. 2014), oxygen (e.g., Neufeld et al. 2010; Flagey et al. 2013; Indriolo et al. 2015), chlorine (e.g., Lis et al. 2010; De Luca et al. 2012; Neufeld et al. 2012, Monje et al. 2013) and fluorine (e.g., Neufeld et al. 2010, Sonnentrucker et al. 2010; this paper) in diffuse molecular clouds and to constrain the physical processes at play in the Galactic diffuse ISM.

 In this work, we focus on two particular hydrides, hydrogen fluoride (HF) and water (H$_2$O) with the aim of further probing the chemistries of fluorine and oxygen-bearing molecules and of determining to what extent both species can be used as diagnostics of the physical processes at play in the diffuse ISM. The combination of the full PRISMAS sample, with some WISH data (Water In Star-forming regions with Herschel; PI: E. F. van Dishoeck) and {\it Herschel} Cycle 1 data allows us to report on the largest set of HF and H$_2$O column density measurements in the Galactic disk to-date. Sections 2 \& 3 describe our set of observations, as well as the reduction and analysis techniques we employed. Section 4 summarizes our results. In Sections 5 \& 6 we compare the HF and H$_2$O distributions measured with our survey to  the most recent 2-sided PDR model predictions (Neufeld \& Wolfire 2009, Hollenbach et al. 2012) for the range of physical conditions most relevant to diffuse molecular clouds. In particular, we validate the role of HF as a tracer of H$_2$ and add weight to the suggestion that H$_2$O can be used as a tracer of H$_2$ in diffuse clouds as well (e.g., Flagey et al. 2013). We confirm the co-spatial distribution of HF and H$_2$O in the Galactic diffuse ISM, which in turn allows us to derive the H$_2$O abundance relative to HF throughout the Galactic disk. In Section 7 we discuss how the measured variations in the H$_2$O abundance relative to HF can be used to test our understanding of the H$_2$O chemistry. We assert the role that grain surface chemistry plays in the production of H$_2$O in the diffuse interstellar medium. Section 8 summarizes our general conclusions. 

\section{Observations}
The names and coordinates of the 12 targets comprising our sample are summarized in Table~\ref{tbl-1}. All targets are well-studied star-forming regions, known to produce strong far-infrared background continuum emission, a prerequisite for detecting absorption by foreground interstellar clouds. The sight lines to each of these sources are also known to intersect multiple spiral arms, hence, allowing us to probe a variety of local physical conditions throughout the Galactic disk (e.g., Godard et al. 2010; Lang et al. 2010).  The sources G$-$0.02$-$0.07 ($+$50 km s$^{-1}$ cloud around SgrA), G$-$0.13$-$0.08 ($+$20 km s$^{-1}$ cloud around SgrA), W28A, W31C, W33A, G34.3$+$0.1, W49N, W51, W3 IRS5 and DR21(OH) were observed through Guaranteed Time Key Program PRISMAS. W3 IRS5 was also observed in Guaranteed time Key Program WISH. W3(OH) and G29.96$-$0.02 were observed as part of the {\it Herschel} Open Time Cycle 1 campaign (OT1, PI: D.A. Neufeld) along with W51 that was reobserved to significantly increase our signal-to-noise (S/N) ratio in the continuum and push our detection limits to lower column density clouds compared to those measured from data previously obtained with the PRISMAS program. 

We observed the ground-state rotational line of HF ($\nu_{\rm rest}=$ 1232.476 GHz; Nolt et al. 1987) in the upper sideband of HIFI band 5a receiver and the ground-state line of para-H$_2$O (p-H$_2$O; $\nu_{\rm rest}=$ 1113.343 GHz)  in the lower sideband of HIFI band 5a receiver. We used multiple Local Oscillator (LO) settings in order to securely identify the HF and p-H$_2$O absorption lines along all 12 sight lines. The proximity of the HF and p-H$_2$O lines in frequency and their detection within the same receiver band ensure that the absorption lines are observed with a very similar telescope performance and that the data are calibrated in a similar fashion. 

We used the Dual Beam Switch (DBS) mode which, combined with the Wide Band Spectrometer (WBS), allows for a spectral resolution of about 1.1 MHz (0.3 km s$^{-1}$ at the HF frequency). The DBS mode uses two reference OFF-beam positions located 3$\arcmin$ on either side of the source position, along an East-West axis. Because the Galactic Center is a very complex and crowded region, we checked for emission or absorption (contamination) in the OFF-beam position using {\it Herschel} observations performed by the HEXGAL project (R. G$\ddot{\hbox{u}}$sten and M. Requena-Torres, private communication). There is no contamination at the OFF-beam position for HF; only minor contamination occurs for p-H$_2$O in the velocity range [$-$ 40, $+$10] km s$^{-1}$. As a result, the absorption components detected in the Galactic disk in this particular velocity range are not included in the analysis described in Section 3 for the two Galactic Center sources G$-$0.02$-$0.07 and G$-$0.13$-$0.08. Contamination in the OFF-beam position is not a concern for the remaining more compact sources in our sample.





\section{Data Reduction and Analysis}
The data were processed to Level 2 with the standard HIFI pipeline in the {\it Herschel} Interactive Processing Environment (HIPE) version 9.1 (Ott 2010), thereby producing fully calibrated spectra for both polarization modes at each LO setting. Further inspection of the Level 2 data showed that the signals obtained in each of the two polarization modes for a given LO setting were in excellent agreement, as were the HF and p-H$_2$O spectra obtained at each LO setting. Occasionally, emission from molecules other than HF or p-H$_2$O appear in the sideband containing the HF or p-H$_2$O absorption (see Fig~\ref{fig1}). For 6 of the 12 sources presented here, Flagey et al. (2013) performed a detailed study of the distribution of water in the ground and excited states accessible to {\it Herschel}/HIFI. While contamination needed to be taken into account for some of the excited states of water, their study showed that the ground-state of p-H$_2$O, of interest here, was mostly free of this effect. For the remaining 6 sources in our sample we  compared the absorption profiles of the ground states of HF and p-H$_2$O across LO settings and for both polarizations. Our comparisons revealed no significant contamination with interloping features for both molecules.  As a result, for each target, we generated average spectra for HF and p-H$_2$O that consist of the weighted sum of up to 6 spectral observations (up to three LO settings with two polarizations each), for which each observation is weighted in inverse proportion to the square of its {\bf root mean square ({\it rms})} noise. The double sideband continuum antenna temperatures T$_A$(cont) and respective {\it rms} noise for HF and p-H$_2$O derived from these weighted average spectra are reported in Table~\ref{tbl-2}, which also lists the total on-source exposure times (t$_{exp}$), the observation dates, and the observation IDs {\bf (Obs IDs)} of the LO settings obtained for HF and p-H$_2$O.  

\subsection{Background Continuum Emission Treatment}

HIFI employs double sideband receivers and for a sideband gain ratio equal to unity, the saturated absorption of radiation at a given frequency for a transition with excitation temperature much less than T$_{\rm A}$(cont) will reduce the measured antenna temperature (T$_{\rm A}$) to one-half the apparent continuum antenna temperature T$_{\rm A}$(cont). 

The top two panels of Figures 1 through 6 display the double sideband weighted average spectra for HF and p-H$_2$O, respectively, {\it versus} Doppler velocity in the Local Standard of Rest frame (V$_{\rm LSR}$) for each of the 12 sight lines probed in our survey.  For most sight lines, HF is detected only in absorption, and the background continuum emission from the source itself is well modeled with a {\bf constant} (red line in each panel). The flux normalized with respect to the continuum flux in a single sideband can then be expressed as, $$\hbox{T$_A$/T$_A$(cont) = (1/2)[1 + $\exp$($- \tau$)]}$$ assuming that the sideband gain ratio is equal to unity. The latter assumption was investigated by Flagey et al. (2013) who found that the HIFI sideband gain ratio in band 5a, the band of interest here, was very stable and indeed consistent with unity. 

\begin{figure}
\plotone{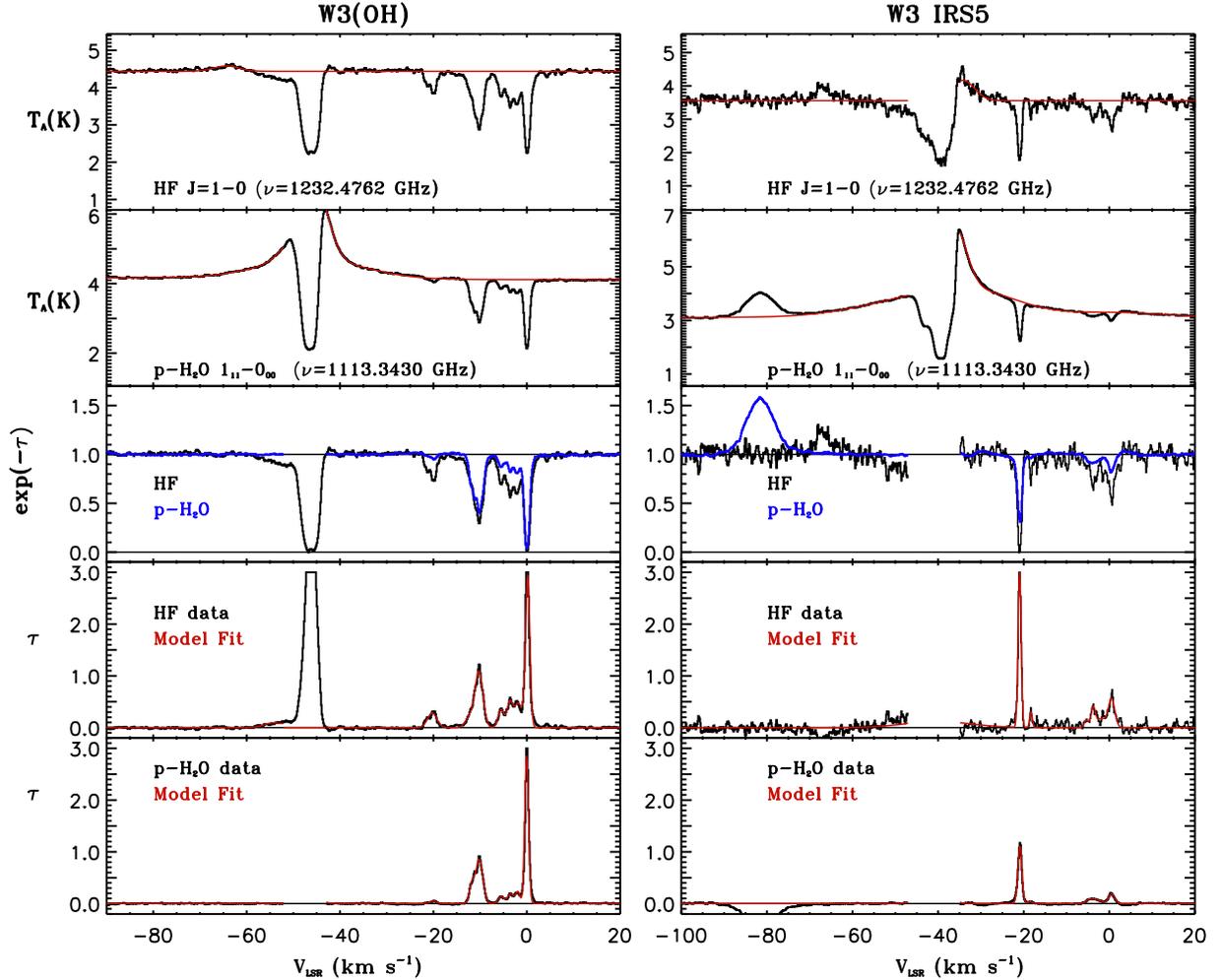}
\caption{Top two panels: Double sideband $rms$-weighted average spectra for HF and p-H$_2$O {\it versus} Doppler velocity in the Local Standard of Rest frame (V$_{\rm LSR}$) toward W3 IRS5 (right) and W3(OH) (left). The best fit model to the background
continuum plus line emission is overlaid in red. The velocity range over which the foreground HF absorption or the foreground p-H$_2$O absorption is blended with emission proximate to the background source is excluded from the modeling. Middle panel: Single sideband continuum normalized spectra for HF (black line) and p-H$_2$O (blue line). The velocity range in which blending occurs is not plotted. Bottom two panels: Optical depth profiles for HF and p-H$_2$O (black lines) versus V$_{\rm LSR}$, respectively.  Overlaid in red are the best multi-Gaussian fit models for the HF and p-H$_2$O profiles for the gas components detected in the foreground of each source that do not suffer saturation. \label{fig1}}
\end{figure}

\begin{figure}
\plotone{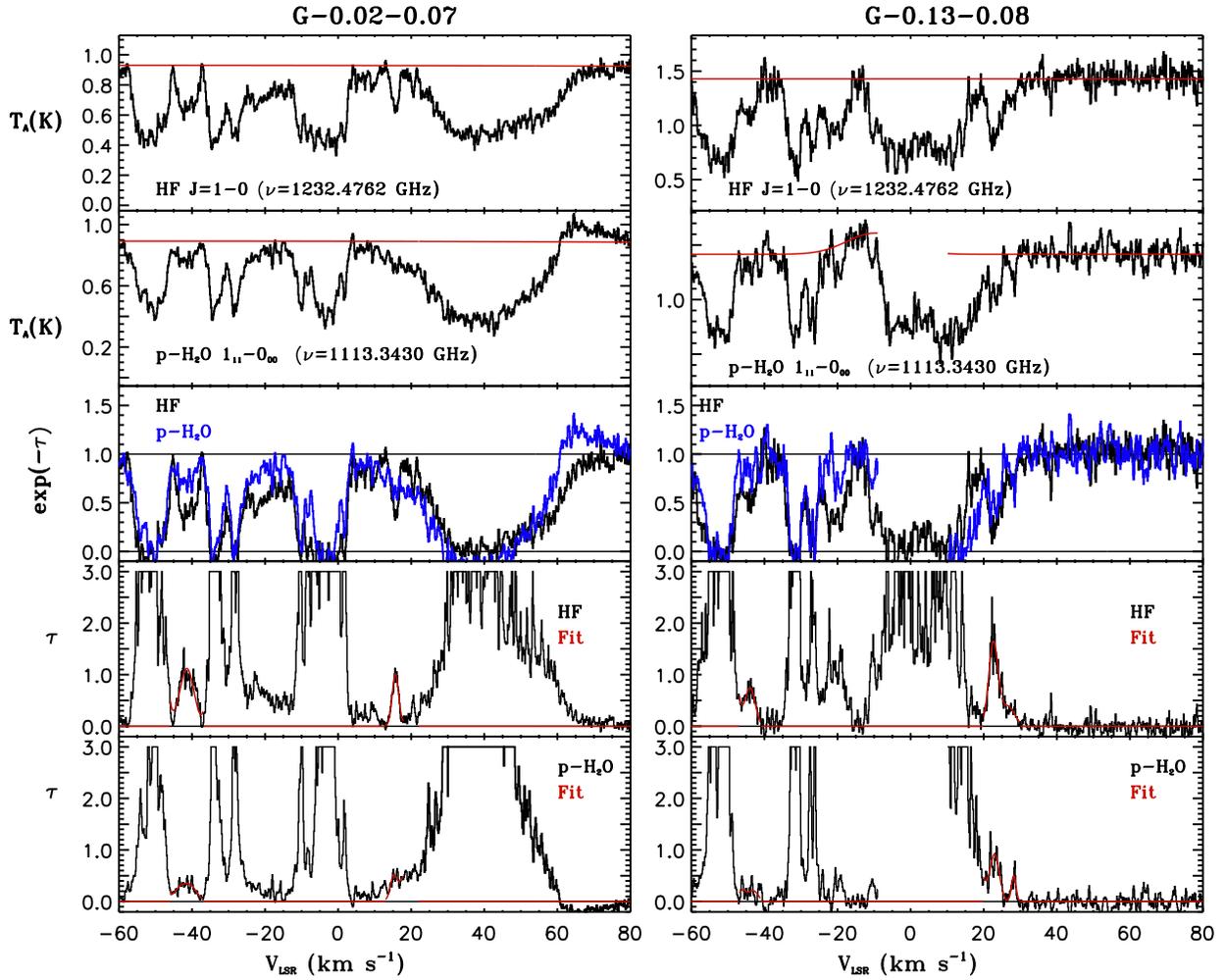}
\caption{Same as in Figure 1 for G$-$0.02$-$0.07 ($+$50 km s$^{-1}$ cloud around SgrA, left) and G$-$0.13$-$0.08 ($+$20 km s$^{-1}$ cloud around SgrA, right). \label{fig2}}
\end{figure}

\begin{figure}
\plotone{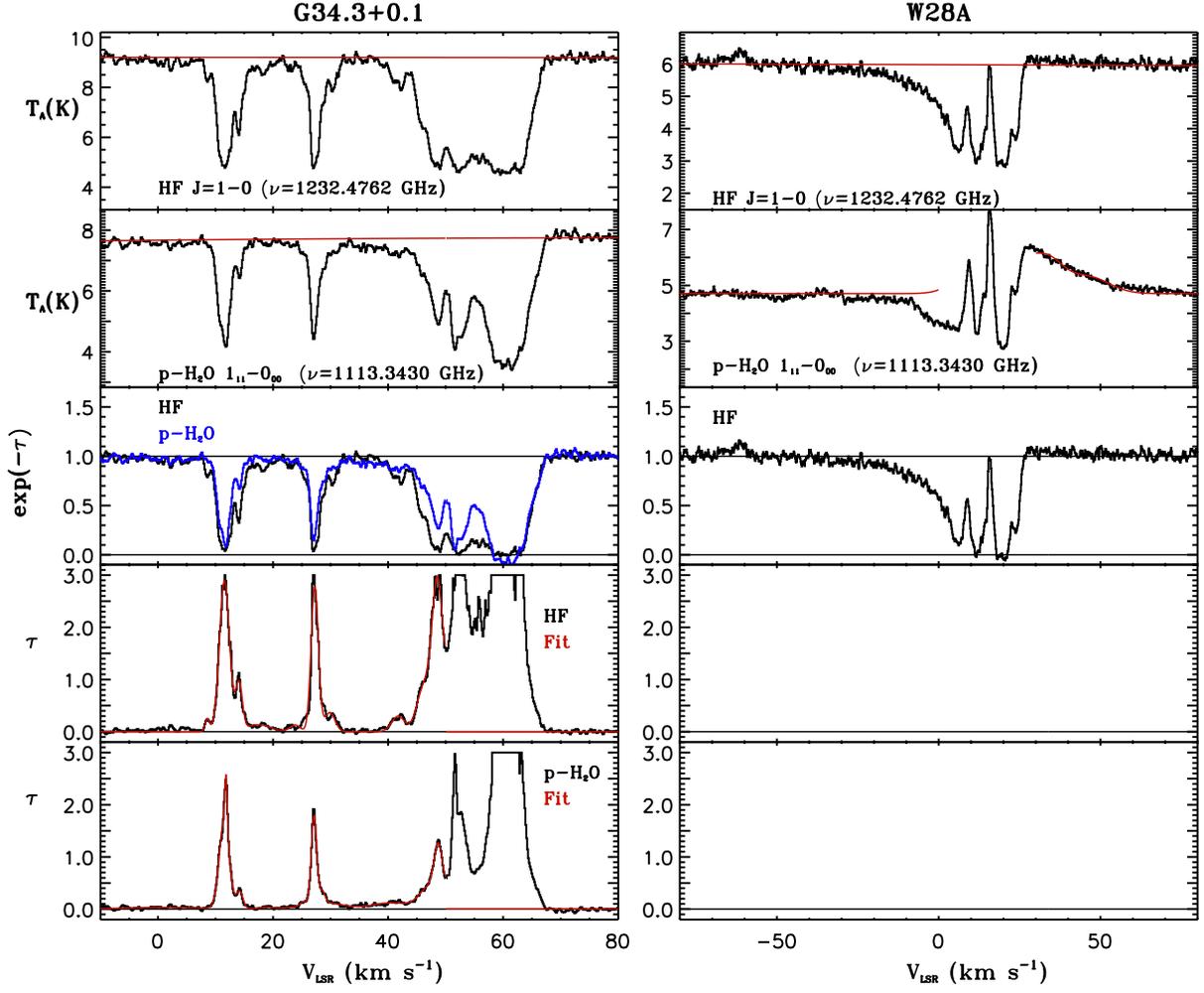}
\caption{Same as in Figure 1 for G34.3$+$0.1 (left). In the case of W28A (right), the velocity ranges of the foreground absorption and the background emission for HF and p-H$_2$O coincide.  As a result, the optical depth profiles and corresponding multi-component Gaussian fits are not presented here. \label{fig3}}
\end{figure}

\begin{figure}
\plotone{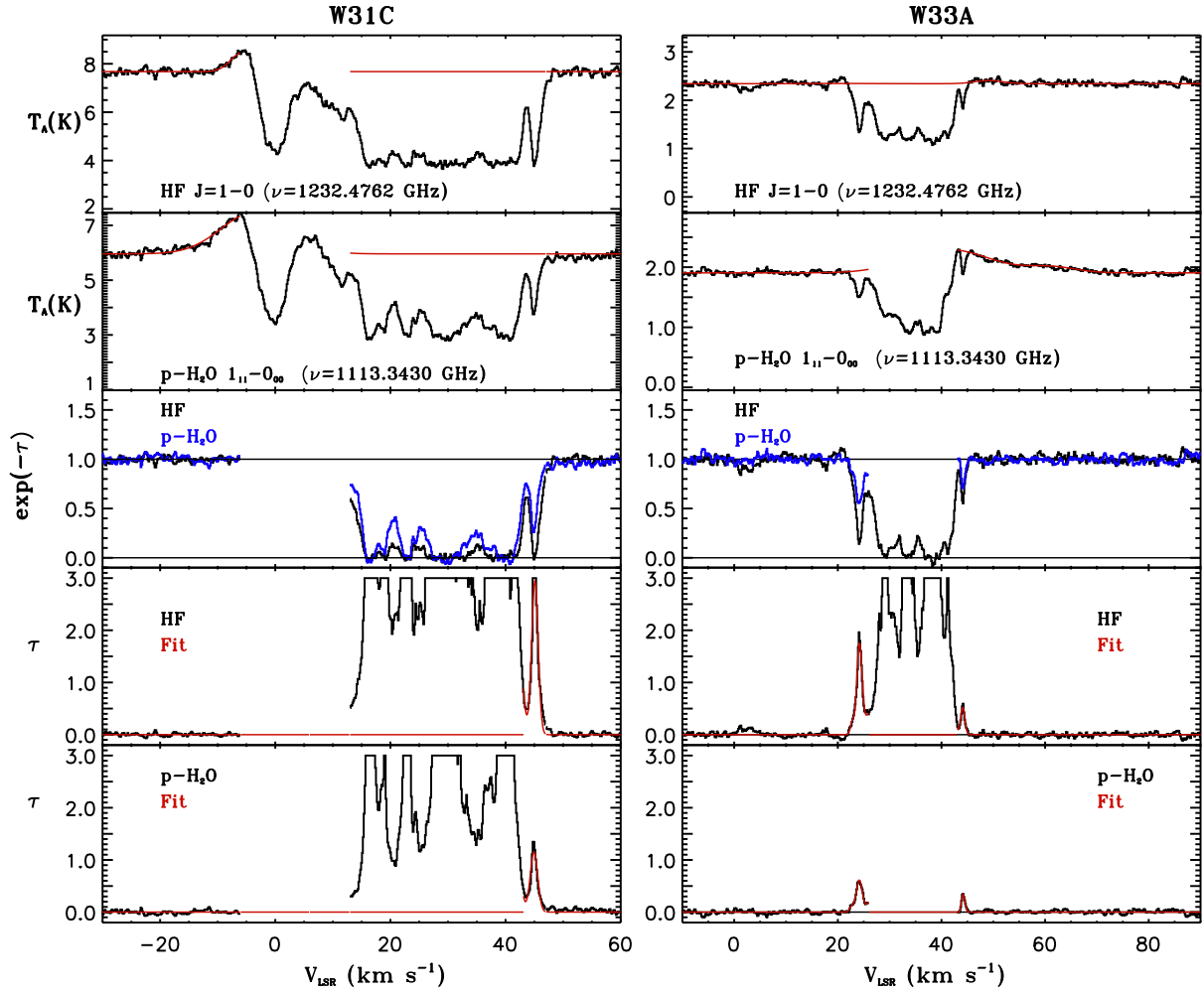}
\caption{Same as in Figure 1 for W31C (left) and W33A (right).\label{fig4}}
\end{figure}

\begin{figure}
\plotone{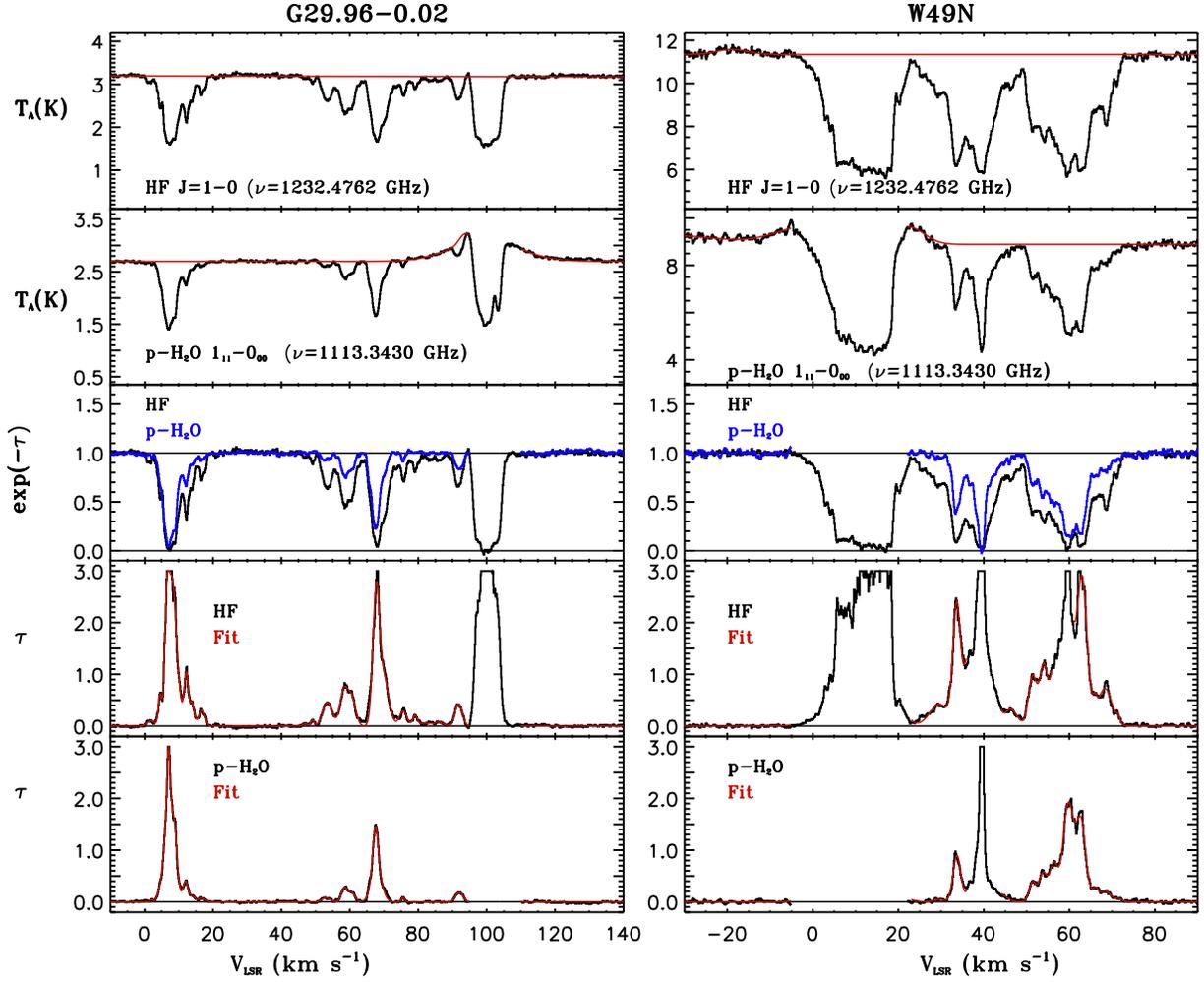}
\caption{Same as in Figure 1 for G29.96$-$0.02 (left) and W49N (right).\label{fig5}}
\end{figure}

\begin{figure}
\plotone{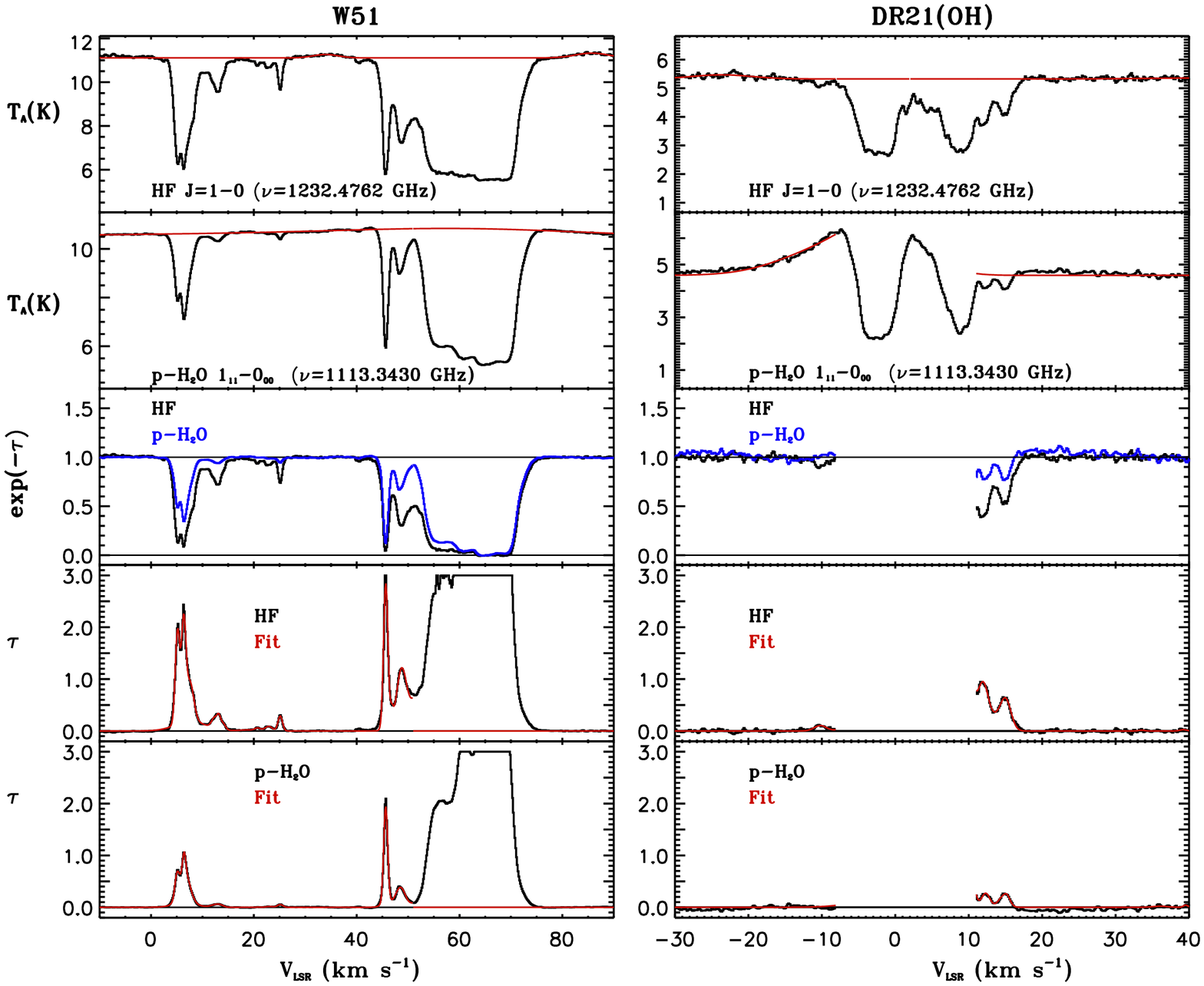}
\caption{Same as in Figure 1 for W51 (left) and DR21(OH) (right).\label{fig6}}
\end{figure}

Toward W31C, W3 IRS5, W28A and W49N, one can see that HF emission arising from gas local to the star-forming region blended itself with HF absorption arising from foreground clouds with projected velocities similar to those of the emitting gas. Such blending also occurs for the p-H$_2$O line absorption toward almost all the sight lines surveyed here. In these cases, we treated the HF and p-H$_2$O emission as features confined to the proximity of the background continuum sources and modeled them as an additional background continuum emission that adds to the dust emission, as was done in Flagey et al. (2013).  For these sight lines, we adopted a linear combination of a zeroth-order polynomial and up to 3 Gaussian profiles to model the shape of the observed background over the regions unaffected by blending with the foreground absorption. In Figs.~\ref{fig1}-\ref{fig6}, we plotted our best fit models in red over the double sideband spectra of HF and p-H$_2$O (black lines) for the velocity regions that can be constrained by our data. Since the H$_2$O emission profiles were constrained using the 1113 GHz features alone, our continuum emission models are not as robust as those obtained by Flagey et al. (2013). As a result, the velocity ranges significantly affected by these continuum emission features are not discussed further here and are not displayed in the bottom two panels of Figs.~\ref{fig1}-\ref{fig6}.

Assuming a sideband gain ratio equal to unity, the flux normalized with respect to the continuum flux in a single sideband is then expressed as $$\hbox{$\exp$($- \tau$)=[T$_A$/T$_A$(cont) $-$ 0.5]/[T$_e$/T$_A$(cont) $+$ 0.5]},$$ where T$_e$ is the continuum emission arising from the HF or p-H$_2$O -containing gas proximate to the star-forming regions and T$_A$ corresponds to the absorption due to the diffuse ISM foreground gas of interest for this study. The middle panels in Figs.~\ref{fig1}-\ref{fig6} display the resulting single sideband, continuum normalized spectrum for HF (black line) and p-H$_2$O (blue line) {\it versus} V$_{\rm LSR}$ for our sample. The horizontal black lines represent the continuum temperature normalized to unity and the zero flux level. One can see that toward all sources, the sideband gain ratios are indeed consistent with unity within our uncertainties (see Table~\ref{tbl-2}).

\subsection{Column Density Measurements}

The HF and p-H$_2$O transitions probed here have spontaneous radiative decay rates of 2.41$\times$10$^{-2}$ s$^{-1}$ and 1.84$\times$10$^{-2}$ s$^{-1}$, respectively. These large rates require high gas densities in order for the collisional de-excitation rate to equal the spontaneous radiative decay rate for both species. The foreground gas we detect through the HF and p-H$_2$O absorptions arises from the Galactic diffuse ISM where gas densities have been measured to be at most $\sim$1000 cm$^{-3}$ (e.g., Jenkins \& Tripp 2001; Sonnentrucker et al. 2007). In the absence of a significant sub-mm radiation field, we expect that these two species will be entirely in their ground rotational states in those foreground gas clouds (e.g., Flagey et al. 2013; Emprechtinger et al. 2013). As a result, the optical depth integrated over velocity for a gas component detected in absorption via the ground state of HF can be written as (see Neufeld et al. 2010)

\noindent  $$\int{\tau dv} = 4.16 \times 10^{-13} N{\rm (HF/cm^2)} \,{\rm km\, s}^{-1}.$$

\noindent Similarly, the optical depth integrated over velocity for a gas component detected in absorption via the ground state of p-H$_2$O  is given by: 

\noindent  $$\int{\tau dv} = 4.30 \times 10^{-13} N{\rm (p-H_2O/cm^2) } \,{\rm km\, s}^{-1}.$$

We used a linear combination of Gaussian components to fit the HF optical depth profiles over the LSR velocity ranges where the profiles showed optically thin or moderately thick HF absorption depths. For frequencies where the optical depth is large (the absorption lines are optically thick), noise fluctuations and small variations in the sideband gain ratio result in large uncertainties of the exact values of the optical depth in each frequency bin. We place a conservative limit of $\tau =$ 3 where (2 T$_A$/T$_A$(cont) - 1) $\leq$ 0.05. For each sight line, the gas distribution was modeled by a linear combination of Gaussian components. The initial number of Gaussian components was determined by eye and compared to the observed spectrum using a $\chi^2$-minimization technique. After each run, one additional Gaussian component was added to the previous model until no significant improvement to the $\chi^2$ value was returned by the minimization algorithm. Given the overall similarity in component distribution between HF and p-H$_2$O, we adopted the best-fit model of the HF spectrum (FWHM and velocity range) as our initial guess to model the p-H$_2$O optical depth profiles for each sight line. We followed the same optimization procedure as for HF to derive our best fit model for the H$_2$O spectra. We found that for one-half of our sight lines, the number of Gaussian components required to best fit  the HF and p-H$_2$O optical depth profiles was identical.  The bottom two panels of Figs.~\ref{fig1}-\ref{fig6} display the HF and p-H$_2$O optical depth profiles (black lines).  For clarity, we do not display the portion of each spectrum where blending between foreground absorption and background emission occurs. Our best fit models are over plotted as red lines only for those absorption components that are optically thin or moderately thick.

While the majority of optically thin or moderately thick absorption features from HF and p-H$_2$O are fitted with a single Gaussian component, some features in our sample require the use of a combination of Gaussian components. In the latter case, we sum the column densities of the individual components constituting the blended complex and report the summed column density over the gas cloud complex. To reflect these differences in sight line structures, Table~\ref{tbl-3} reports the LSR velocity range over which the optical depth absorptions were integrated rather than the individual Gaussian parameters resulting from modeling the sight lines. We only report and discuss measurements for those gas components that exhibit optically thin or moderately thick profiles ($\tau <$ 3) for both HF and H$_2$O.




\section{Results}

The spectra displayed in Figs.~\ref{fig1}-\ref{fig6} show that HF is detected in all gas clouds already known to trace the diffuse ISM based on HI  21 cm observations (e.g., Lang et al. 2010), or on HCO$^+$ absorption surveys (e.g., Godard et al. 2010). Our survey hence demonstrates that the diffuse molecular phase as traced by HF is as ubiquitous as the diffuse atomic phase in the Galactic spiral arms. Our survey also clearly establishes that water is as widespread as HF or H$_2$ in this phase. Toward the Galactic disk sources, we find that the water velocity distribution resulting from our sight line modeling is identical to that of HF, within 1 km s$^{-1}$, clearly indicating that the water distribution traces that of HF almost perfectly once $N$(H$_2$O) is in excess of 10$^{12}$ cm$^{-2}$. This striking similarity was noted in earlier work (Neufeld et al. 2010; Sonnentrucker et al. 2010) and is now confirmed throughout the Galactic volume we probed. One exception is the gas localized in the immediate vicinity of the Galactic Center where the HF distribution is observed to be confined to more discrete gas features than the water distribution (e.g., Lis et al. 2010; Monje et al. 2011, Sonnentrucker et al. 2013). For the Galactic disk sources in common between Flagey et al. (2013) and this paper, the velocity coincidence of the water absorption features with the extent of the continuum emission due to material close to the source led Flagey et al. (2013) to point out that the water absorption features detected toward these sources might not necessarily originate in Galactic disk gas. When combining the remarkable similarities in the HF and water distributions with the lack of evidence for significant HF emission arising from gas local to the background sources, our analysis indicates that the HF and water features we considered are mostly tracing the foreground Galactic disk material. Our conclusions are consistent with those derived in a recent Galactic [C{\rm II}] survey by Gerin et al. (2015) who showed that the C$^+$ absorption behaves like a foreground screen to the background sub-mm continuum sources.

Table~\ref{tbl-3} summarizes our line of sight modeling results and reports on our column density measurements as follows. Column 1 lists the LSR velocity range over which the optical depth profiles for HF and p-H$_2$O were integrated. Columns 2 and 3 report our column density measurements for HF, $N$(HF),  and para-water, $N$(p-H$_2$O) with corresponding  1$\sigma$ uncertainties in units of 10$^{12}$ cm$^{-2}$. In Column 4 we list the total column density of water, $N$(H$_2$O)$_{\rm tot}$, using an ortho-to-para ratio of 3  for water as measured by Flagey et al. (2013) for the Galactic disk. We note that Lis et al. (2010, 2013) measured an ortho-to-para ratio around 2.35 for water toward the Galactic Center source Sgr B2(M). This is the only value below the LTE limit that has been found so far and the effect on the H$_2$O column density is rather minor ($\sim$20\% decrease).  Finally, Column 5 reports the abundance of H$_2$O relative to HF,  $N$(H$_2$O)$_{\rm tot}$/$N$(HF) and associated 1$\sigma$ uncertainty.

We report column density {\it measurements} for both HF and p-H$_2$O for a total of 47 absorption features, all sight lines considered, and {\it limits} on either the p-H$_2$O or the HF column densities for 5 gas components. Thirty of these 47 gas features are satisfactorily modeled with a single Gaussian component within our resolution (0.3 km s$^{-1}$ at the HF frequency) and exhibit FWHM ranging from $\sim$1.0 to 4.3 km s$^{-1}$. The remaining 17 absorption complexes have  "effective" FWHM varying from $\sim$4.5 to 10 km s$^{-1}$. We also find that the FWHM derived from the p-H$_2$O profile modeling are identical to those for HF within 1 km s$^{-1}$, a similarity already noticed in previous work (Neufeld et al. 2010; Sonnentrucker et al. 2010) and readily seen in Figs.~\ref{fig1}-\ref{fig6} in this work. Note that this similarity between the HF and H$_2$O distributions is genuine throughout the Galactic disk and does not result from our line of sight modeling strategy since HF and H$_2$O are fitted separately. Our survey therefore shows that HF and H$_2$O trace, in general, the same gas volumes throughout  the Galactic disk. In the absence of H$_2$ or CH data, one can thus consider using HF to trace H$_2$ in gas components exhibiting H$_2$O absorption,  as will be discussed in the next sections.

\begin{figure}
\epsscale{.9}
\plotone{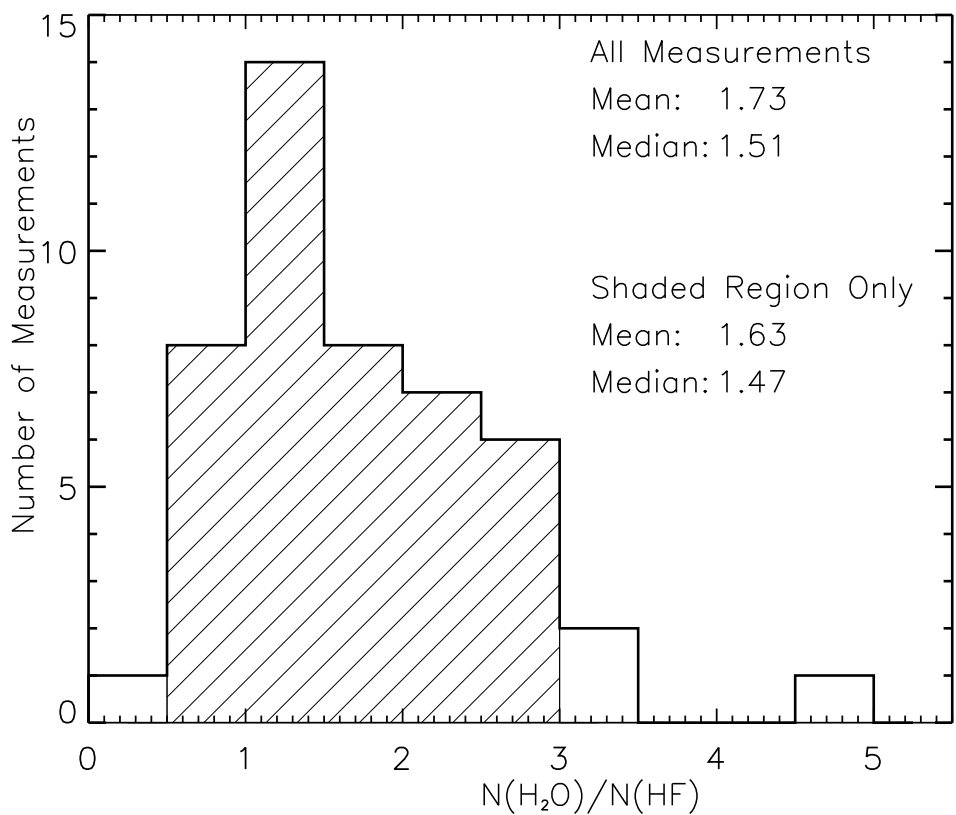}\caption{Distribution of the H$_2$O abundance relative to HF in the Galactic disk gas that we sampled. The black-contour histogram displays the distribution using all 47 measurements. The mean and median of this distribution are given in the top right corner.  Four ratios differ from the mean by more than a factor of 2. When excluding those outliers (gray-shaded histogram), the distribution of the H$_2$O abundance relative to HF yields the mean and median values reported in the middle right. \label{fig11}}
\end{figure}

Figure~\ref{fig11}  displays the histogram distribution of the $N$(H$_2$O)/$N$(HF)  ratio for the 47 features that yielded column density measurements for both HF and H$_2$O. With this sample, we derive a mean water abundance relative to HF of 1.73 with a standard deviation around the mean of 0.87. Four velocity ranges exhibit ratios that differ from the mean value by more than a factor of 2. When excluding the latter outliers from the distribution (gray-shaded histogram), we derive a mean H$_2$O to HF abundance ratio of 1.63 with a standard deviation around the mean of 0.58.  The small differences in the mean values for both distributions (less than 12\%)  indicate that the water abundance distribution relative to HF is remarkably narrow throughout the Galactic disk and quite constant to within a factor of 2. Considering the non-Gaussian nature of the distribution, we adopted the median value of the full (47-point) distribution as our best measurement of the water abundance relative to HF,  $N$(H$_2$O)/$N$(HF)$=$1.51, for the Galactic disk. The relative constancy of the latter ratio (to within a factor 2.5) was noted earlier by Flagey et al. (2013) for a smaller number of gas features. Their measured water abundance relative to H$_2$ of 5$\times$10$^{-8}$ leads to an estimate of the  H$_2$O/HF ratio of $\sim$1.4, in agreement with our measurements within uncertainties. Our survey therefore extends the conclusions of Flagey et al. to a larger volume in the Galactic disk. We note that Flagey et al. (2013) used HF/H$_2$$=$3.6$\times$10$^{-8}$ which will tend to underestimate the H$_2$ column density and overestimate H$_2$O abundance relative to H$_2$. We also note that the deviations around the median value for the H$_2$O-to-HF ratio indicate that true variations in the H$_2$O/HF ratio do exist. In the following sections, we will argue that these variations are mostly due to local variations in the H$_2$O/H$_2$ ratio rather than variations in the HF/H$_2$ ratio in the ISM.  As a result, our measurements offer a unique opportunity to test the physical processes as well as the chemical pathways involved in the production of gas-phase H$_2$O in the diffuse interstellar medium.


\section{HF Abundance in the Galactic disk}

In the diffuse ISM, the defining tracers of the atomic and molecular phases are atomic hydrogen (H$^0$) and molecular hydrogen (H$_2$).  Direct measurement of H$_2$ are most easily obtained through its ground electronic transitions which lie in the far-UV shortward of 1150\AA\ in the Galaxy. As a result, significant effort has been dedicated for decades to identifying surrogate tracers of H$_2$ in the various phases of the ISM.  For diffuse molecular cloud environments, combinations of FUV and optical observations have demonstrated that CH is a valuable tracer of H$_2$. The observed CH-H$_2$ relationship is routinely used to estimate H$_2$ column densities in the absence of H$_2$ observations. This relationship, however,  exhibits a large but real dispersion of 0.2 dex around its mean value of 3.5$\times$10$^{-8}$ (Sheffer et al. 2008, and references therein; Levrier et al. 2012). The H$_2$ column densities derived from CH measurements are, thus, only accurate to within a factor of $\sim$2. The range in H$_2$ column densities traced by CH is also limited as the weak CH absorption is typically below detection level for diffuse molecular clouds with $N$(H$_2$) $\le$${\rm few} \times$ 10$^{19}$ cm$^{-2}$. CH depletion onto dust grains, or ``freeze-out'' limits the use of CH as a probe of H$_2$ in denser clouds where $N$(H$_2$) $\ge$ 3$\times$ 10$^{22}$ cm$^{-2}$ (Mattila 1986).  

Theoretical models of interstellar chemistry have predicted that HF will be the dominant reservoir of gas-phase fluorine over a large range of physical conditions (Neufeld et al. 2005; Neufeld \& Wolfire 2009). In particular, HF  is expected to trace H$_2$ both in diffuse gas with molecular fractions of only a few percent ($N$(H$_2$) $\ge$${\rm few} \times$ 10$^{18}$ cm$^{-2}$, where CH is below the current detection limit) and in dense clouds of larger molecular fractions where HF ``freeze-out''  is not yet significant. As a result, HF is predicted to be the most sensitive tracer of H$_2$  in the diffuse molecular regime probed here. One of the goals of the PRISMAS key program was to test the latter predictions and determine whether HF could be used as a surrogate tracer of H$_2$ in the sub-mm range where H$_2$ is not observable directly.  In the next sections we describe the modifications applied to our diffuse cloud models and the updates made to the chemical networks used in the models.  We compare the HF observations currently available against those new model predictions and we discuss our understanding of fluorine chemistry in the diffuse ISM.

\subsection{Model Modifications and Predictions}

 We have used a modified version of the diffuse cloud models presented in \cite{Neufeld2005} and \cite{Neufeld2009}. The modifications include ice freeze out and  grain-surface chemistry as in \cite{Hollenbach2009}, 
the gas-phase reactions and radiation field attenuation as in \cite{Wolfire2010}, and the oxygen chemistry as in \cite{Hollenbach2012}.
 The model consists of a slab of gas of constant hydrogen nucleus 
 density $n_{\rm H}$, and total width $A_{V,{\rm tot}}$, which is 
  illuminated by the interstellar radiation field, $\chi$, measured
  in units of the \cite{Draine1978} field.
  The gas temperature and the abundances of atomic and molecular species are
  calculated as a function of $A_V$ through the cloud 
   under the assumptions of thermal balance, and chemical equilibrium. We 
    assume the slab is embedded in an isotropic interstellar radiation field 
  of value $\chi$ in free-space and thus $\chi/2$ is incident on opposite sides of the slab.
  We use the single ray approximation given in \cite{Wolfire2010} for attenuation of the
    isotropic 
  field. 

\begin{figure}[hf]
\epsscale{0.5}
\plotone{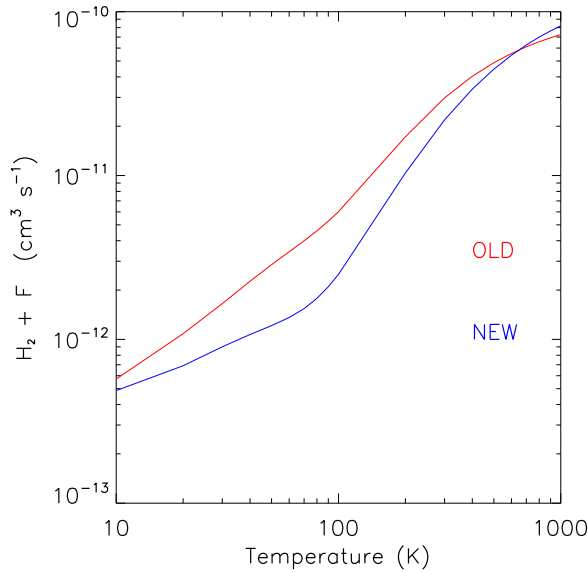}
\caption{Reaction rates for HF production
using the fitted functions given in \cite{Neufeld2005} (old rate) and \cite{Tizniti2014} 
(new rate). Both assume ${\rm H_2}$ in LTE. Over typical cloud temperatures in the diffuse
ISM, the \cite{Tizniti2014} rate is a factor $\sim 2.5$ lower than the previous rate resulting in lower HF abundances.\label{fig7}}
\end{figure}

  An important update from the \cite{Neufeld2009} 
  paper is a new measurement for the rate of the reaction
  $${\rm F}\, +\, {\rm H_2}\rightarrow {\rm HF}\, + \, {\rm H}$$
   \citep{Tizniti2014}.  This reaction is the dominant formation 
   route for HF in regions with both high or low molecular fraction. 
   The previously used rate was based on a fit to the calculations 
   of \cite{Zhu2002} for an assumed ${\rm H_2}$ population in local
   thermodynamic equilibrium (LTE). The new rate is based on experimental
   measurements between 11 K and 295 K using a supersonic flow technique. 
   \cite{Tizniti2014} provide a fit to the rate coefficient {\it versus} temperature
   assuming ${\rm H_2}$ in LTE. Figure~\ref{fig7} compares the new and old rate using
  the fitted functions.  Over temperatures expected in diffuse gas $T \sim 50-100$ K, the new
   rate is a factor of $\sim 2.5$ times lower than the old rate and thus at low 
  ${\rm H_2}$ abundance, the new rate results
   in a lower HF abundance by a factor of $\sim 2.5$. The HF abundance
  increases with ${\rm H_2}$ abundance until all of the gas phase F 
  is locked in HF, at which point  the HF abundance is insensitive to
  the ${\rm H_2\, +\, F}$ reaction rate.

 Additional updates to the previous codes 
  include fine-structure collision rates for O~\small{I}  and C~\small{I} with H$^0$ from 
   \cite{Abrahamsson2007}, and revised photo rates with PAHs using the
  \cite{Draine1978} radiation field with linear yield functions for the
  ionization of PAH$^0$ and electron detachment of PAH$^-$.
Finally, an important parameter for the gas-phase production of
  water is the cosmic-ray ionization rate, $\zeta$. We use the definition
  that $\zeta$ is the primary rate per hydrogen nucleus. 
  The ionization rate is often quoted as the total rate per ${\rm H_2}$ which is 
  a factor $\sim 2.3$ times larger in molecular gas.
 
We have run a series of models varying the total extinction through the cloud $A_{V,{\rm tot}}$, the
density $n_{\rm H}$, the cosmic-ray ionization rate $\zeta$, and the incident radiation field $\chi$.
The parameters are intended to cover a range of values expected in diffuse molecular clouds
\citep[e.g.,][]{Sonnentrucker2007, Phillips2013, Goldsmith(2013), Indriolo2013}.
The cloud extinction varies between $A_{V,{\rm tot}}\sim 0.05$ mag (depending on other
parameters) and  $A_{V,{\rm tot}} =  4$ mag. The density varies between $n_{\rm H} = 50$ ${\rm cm^{-3}}$ and
$n_{\rm H} = 900$ ${\rm cm^{-3}}$. As a standard model we use a radiation field of $\chi=1$, but
examine the case $\chi=3$ as expected for the field in the inner Galaxy \citep{Wolfire2003}.
We adopt a standard cosmic-ray ionization rate of $\zeta=2\times 10^{-16}$ s$^{-1}$. This rate is close 
to the average rate of $\sim 1.5\times 10^{-16}$ ${\rm s^{-1}}$
estimated from ${\rm H_3^+}$ column density measurements over 50 lines-of-sight
by Indriolo \& McCall (2012b). It is also close to the rate of $\sim 2.1\times 10^{-16}$ ${\rm s^{-1}}$
estimated by Indriolo et al. (2012a) in the sight line towards W51. Previous diffuse ISM studies adopted cosmic ray ionization rates which are lower than the standard value we adopted by a factor of $\sim$10. We investigate the effects of lower rates on our predictions as well. For the gas-phase abundances of carbon and oxygen we adopt the values of 1.6$\times 10^{-4}$ \citep{Sofia2004} and 3.9$\times 10^{-4}$
\citep{Cartledge2004}, respectively. For fluorine,  we use a gas-phase abundance
value of 1.8$\times 10^{-8}$ from \cite{Neufeld2005} which amounts to
$\sim 60$\% of the solar abundance with the remainder depleted onto grains.
Our results for the HF abundance will scale linearly with the adopted gas-phase abundance of fluorine.

\begin{figure}
\epsscale{1.}
\plotone{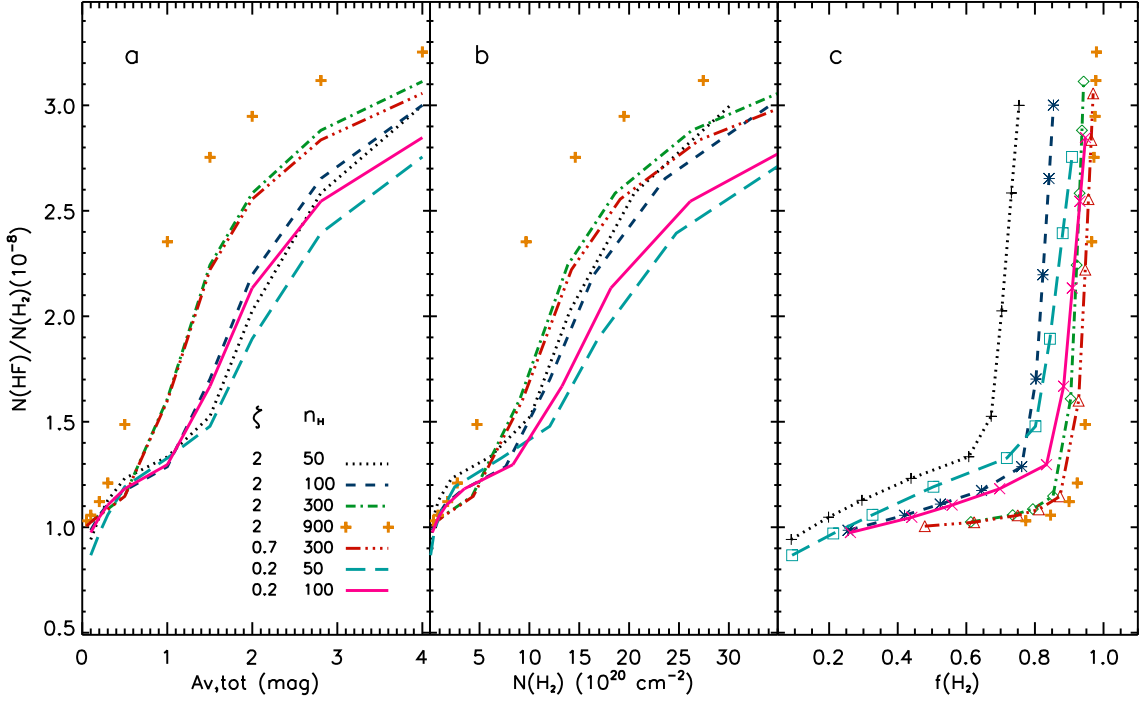}
\caption{Predictions for the HF abundance relative to H$_2$ in a single cloud as a 
function of total visual extinction ($A_{V,{\rm tot}}$, panel a), total H$_2$ column density ($N$(H$_2$), panel b), and molecular fraction ($f({\rm H_2})$, panel c) using our modified 2-sided PDR model. The UV field is given in units of the Draine (1978) field, is fixed to $\chi=$1 and is assumed to be incident isotropically at a value of $\chi=$ 1/2 on each side of the cloud. In panel a), we list the range in cosmic-ray ionization rate $\zeta$ (10$^{-16}$ s$^{-1}$) and density $n_{\rm H}$ (${\rm cm^{-3}}$) we explored. In panel c), colored symbols show model predictions for clouds with $A_{V,{\rm tot}}=$ 0.1, 0.2, 0.3, 0.5, 1.0, 1.5, 2.0, 2.8 and 4.0 mag. For models ($n_{\rm H}=$ 300 ${\rm cm^{-3}}$ and $\zeta=$ 0.7$\times$10$^{-16}$ s$^{-1}$) or ($n_{\rm H}=$ 900 ${\rm cm^{-3}}$ and  $\zeta=$ 2$\times$10$^{-16}$ s$^{-1}$), $A_{V,{\rm tot}}=$ 0.05 mag is also shown. To convert $A_{V,{\rm tot}}$ to hydrogen nuclei column density $N_{\rm H}=$ $N$(H$^0$) + 2$N$(H$_2$) we use $N_{\rm H}$(cm$^{-2}$)$=$ 2$\times$10$^{21}$ $A_{V,{\rm tot}}$. 
\label{fig8}}
\end{figure}

Figure~\ref{fig8} shows our model predictions for the HF column density  relative to the H$_2$
column density  
as a function of total visual extinction in the cloud $A_{V,{\rm tot}}$ (panel a), as a function of the total ${\rm H_2}$
column in the cloud $N({\rm H_2})$ (panel b), and as a function of the ${\rm H_2}$ fraction in the cloud
$f({\rm H_2})=2N({\rm H_2})/[N({\rm H^0})+2N({\rm H_2})]$ (panel c). For the models shown in Fig~\ref{fig8}, $\chi$ is fixed at  $\chi=$1 in units of the Draine (1978) field. The $N({\rm HF})/N({\rm H_2})$ ratio is predicted to
rise monotonically with increasing $A_{V,{\rm tot}}$,  $N({\rm H_2)}$, and $f({\rm H_2})$ and varies
by at most a factor of $\sim 3$ while $\zeta$ and $n_{\rm H}$ vary by at least a factor of 10 over the entire parameter range considered here. 

The dependence of the $N$(HF)/$N$(H$_2$) ratio on total cloud extinction $A_{V,{\rm tot}}$ can be understood by considering the dominant formation and destruction processes
for HF at the cloud center. The detailed  ${\rm HF}$ chemistry is presented 
in \cite{Neufeld2005} and \cite{Neufeld2009}. Briefly, the production of HF proceeds by the reaction of ${\rm H_2}$ with
F, while the destruction occurs through reactions with ${\rm C^+}$, H$_3$$^+$, He$^+$, and Si$^+$ plus
photodissociation.
For the low density models, ${\rm C^+}$ 
dominates the destruction of HF, to $A_{V,{\rm tot}}\sim 1.5$ mag, at which
point electrons recombining with ${\rm C^+}$ lead to a lower
${\rm C^+}$ abundance and higher HF abundance. At high density
($n_{\rm H} \ge 300$ ${\rm cm^{-3}}$), the recombination of ${\rm C^+}$
is drawn to the cloud surface leading to higher HF columns even
for clouds of low $A_{V,{\rm tot}}$. Increasing cloud column densities results in a drop in the C$^+$ abundance and reactions with less abundant ions dominate the destruction
of HF. Thus the HF destruction rates fall and the $N$(HF)/$N$(H$_2$) ratio rises.

Since both the HF column density and ${\rm H_2}$ column density depend
on the ${\rm H_2}$ abundance, the ratio does not depend directly
on the ${\rm H_2}$ abundance. In addition,  
since both the formation of HF and destruction of HF have
the same dependence on density ($\propto n^2$),
the HF fractional abundance is not directly density
dependent. As a result, we only see a weak density dependence in
Figure~\ref{fig8}. 

The HF abundance predictions are also mostly insensitive to variations in $\zeta$. Note that a slight increase of the HF abundance of at most 10\% is predicted at  $A_{V,{\rm tot}}$= 4 mag for $n_{\rm H}$ $\leq$ 100 cm$^{-3}$ when increasing $\zeta$ by a factor of 10. Cosmic rays play a role in the destruction of both HF and H$_2$, as they produce H$_3$$^+$ and He$^+$ which then react with HF and H$_2$.  The decrease in the H$_2$ abundance is greater than that in the HF abundance which, in turn, results in the $\sim$10\% increase in the $N$(HF)/$N$(H$_2$) ratio seen in Fig.~\ref{fig8}.

\subsection{HF Observations}

The grid of model calculations presented in the previous section predicts that the column density of HF with 
respect to H$_2$ varies between $N({\rm HF})/N({\rm H_2})$ $\sim$ 0.9 $\times 10^{-8}$ at the cloud surface ($A_{V,{\rm tot}}\sim$ 0.08 mag) in the low density regime to $N({\rm HF})/N({\rm H_2}) \sim 3.3 \times 10^{-8}$ at the cloud center ($A_{V,{\rm tot}}\sim$ 4 mag) in the high density regime considered here. As a result, HF is predicted to be a very sensitive probe of H$_2$, as it can trace H$_2$ both in mostly diffuse atomic gas ($f$(H$_2$)$\leq$ 0.1) and in mostly diffuse molecular gas (0.1$<$ $f$(H$_2$) $\leq$0.9) in the interstellar medium. The HF abundance in our models never reaches the value of  3.6 $\times$ 10$^{-8}$ expected if all gas-phase F is locked into gas-phase HF; a value derived from FUV measurements of the abundance of gas-phase atomic fluorine (e.g., Snow et al. 2007). The leveling off around a value of 3.3 $\times 10^{-8}$ seen in Fig.~\ref{fig8}  for $n_{\rm H}$ = 900 cm$^{-3}$, is due to the incomplete conversion of F into HF in the outer regions of the cloud. If the gas becomes fully molecular slightly before all F is locked in HF, then the total column density ratio $N$(HF)/$N$(H$_2$) is less than the F abundance. 

Following the first {\it Herschel}/HIFI detections of the ground rotational transition of HF toward sight lines comprised mostly of diffuse atomic/molecular clouds (Neufeld et al. 2010; Sonnentrucker et al. 2010), numerous additional HF detections were reported both with the high resolution HIFI data and the lower resolution SPIRE and PACS data (e.g., Kirk et al. 2010). At this point in time, HF has been detected in the Galactic Center (e.g., Sonnentrucker et al. 2013; Goicoechea et al. 2013), in hot cores around massive proto-stars (e.g., Phillips et al. 2010; Emprechtinger et al. 2012), in the ISM of extra-galactic sources up to z=2.6 (e.g., Monje et al. 2011; Monje et al. 2014) and in the Galactic disk (Flagey et al. 2013; this work). These studies clearly established the ubiquitous presence of the HF molecule in the interstellar medium, as predicted by our models and as expected from a tracer of H$_2$.

The $N$(HF)/$N$(H$_2$) values currently available in the literature range from 1.0 $\times$ 10$^{-8}$ to  2.5$\times$ 10$^{-8}$ (e.g. Sonnentrucker et al. 2010; Phillips et al. 2010, Monje et al. 2011; Emprechtinger et al. 2012) and are fully consistent with our model predictions. With the exception of the HF/H$_2$ ratio estimated by Indriolo et al. (2013) and discussed below, those ratios were estimated {\it indirectly} from simultaneous measurements of HF and either CH or $^{13}$CO along these sight lines. The CH column densities were converted into H$_2$ columns using the most recent estimate of the CH-H$_2$ relationship $N$(CH)/$N$(H$_2$)=3.6 $\times$ 10$^{-8}$ (Sheffer et al. 2008). As mentioned earlier, the CH-H$_2$ relationship exhibits a standard deviation of about 0.2 dex that is real and thought to be related to the chemical pathways involved in the formation of CH in the diffuse ISM. The $^{13}$CO column densities where converted to H$_2$ columns for a given carbon Galactic isotopic ratio and for a given CO-to-H$_2$ ratio; the latter two quantities also vary by factors of a {\rm few} in the Galactic disk (Sonnentrucker et al. 2007; Langer \& Penzias 1990). With these caveats in mind, it is interesting to note that the range in the $N$(HF)/$N$(H$_2$) ratio measured to-date with {\it Herschel} is fully consistent with our model calculations, thus, validating our understanding of both the physical conditions and the new reaction rates involved in the chemistry of interstellar fluorine.

A few {\it direct} measurements of $N$(HF)/$N$(H$_2$) were recently obtained by Indriolo et al. (2013) for a handful of targets sampling diffuse molecular clouds where $N$(H$_2$) and $N$(HF) were derived from FUV and near-IR observations, respectively. For all sight lines where measurements of both quantities were obtained, Indriolo et al. (2013) derived $N$(HF)/$N$(H$_2$) ratios between 0.5 and 1.4 $\times$ 10$^{-8}$. Of particular interest are the results obtained toward the translucent sight line HD154368. This sight line was best modeled to be comprised of diffuse molecular material with $n_{\rm H}$= 325 cm$^{-3}$, $\chi$=3 and $\zeta$= 0.5$\times$ 10$^{-16}$ s$^{-1}$ by Spaans et al. (1998) while FUV measurements based on C$_2$ observations yielded $n_{\rm H}$= 240 cm$^{-3}$ (Sonnentrucker et al. 2007). Our model calculations for the HF abundance using the parameters from Spaans et al. (1998) and an average density of $n_{\rm H}$ = 300 cm$^{-3}$ yield a ratio of $N$(HF)/$N$(H$_2$) = 2.14 $\times$ 10$^{-8}$ toward this sight line. Our model predictions are consistent with the direct measurements of Indriolo et al. (2013) of  $N$(HF)/$N$(H$_2$) = 1.15 $\pm$ 0.41 within 3-$\sigma$. In summary, all measurements obtained to date -whether direct or indirect- corroborate the prediction that HF can be used as a surrogate tracer of H$_2$ in the diffuse ISM, down to very low molecular fractions where CH is typically below our detection level. Since direct H$_2$ measurements are not available for the entire set of HF and H$_2$O absorption features we report in this survey,  we will use the HF column densities we derive as proxies for H$_2$ in the remainder of the paper.

\section{H$_2$O Abundance in the Galactic disk}

\subsection{Model Predictions}

\begin{figure}
\epsscale{1.0}
\plotone{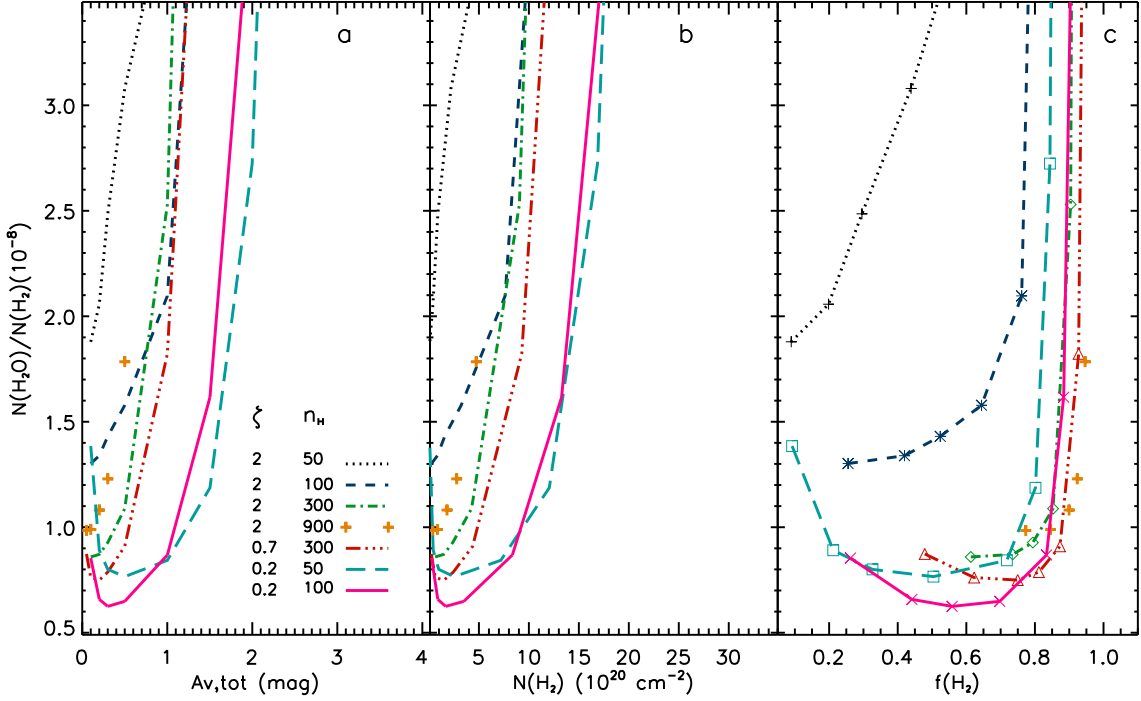}
\caption{Predictions for the H$_2$O abundance relative to the H$_2$ in a single cloud as a 
function of total visual extinction ($A_{V,{\rm tot}}$, panel a), total H$_2$ column density ($N$(H$_2$), panel b), and molecular fraction ($f({\rm H_2})$, panel c) using our modified 2-sided PDR model. The UV field is given in units of the Draine (1978) field, is fixed to $\chi=$1 and is assumed to be incident isotropically at a value of $\chi=$ 1/2 on each side of the cloud. In panel a), we list the range in cosmic-ray ionization rate $\zeta$ (10$^{-16}$ s$^{-1}$) and density $n_{\rm H}$ (${\rm cm^{-3}}$) that we explored. In panel c), colored symbols show model predictions for clouds with $A_{V,{\rm tot}}=$ 0.1, 0.2, 0.3, 0.5, 1.0, 1.5, 2.0, 2.8 and 4.0 mag. For models ($n_{\rm H}=$ 300 ${\rm cm^{-3}}$ and $\zeta=$ 0.7$\times$10$^{-16}$ s$^{-1}$) or ($n_{\rm H}=$ 900 ${\rm cm^{-3}}$ and  $\zeta=$ 2$\times$10$^{-16}$ s$^{-1}$), $A_{V,{\rm tot}}=$ 0.05 mag is also shown. To convert $A_{V,{\rm tot}}$ to hydrogen nuclei column density $N_{\rm H}=$ $N$(H$^0$) + 2$N$(H$_2$) we use $N_{\rm H}$(cm$^{-2}$)$=$ 2$\times$10$^{21}$ $A_{V,{\rm tot}}$.
\label{fig9}}
\end{figure}

Figure~\ref{fig9} shows the predictions of the PDR model described in the previous section for the ${\rm H_2O}$
column density  relative to the H$_2$
column density  
as a function of total visual extinction in the cloud ($A_{V,{\rm tot}}$, panel a), 
as a function of the total ${\rm H_2}$ column density in the 
cloud ($N({\rm H_2})$, panel b)  and as a function of the molecular fraction
 in the cloud ($f({\rm H_2})$, panel c). The UV field is given in units of the Draine (1978) field and is fixed to $\chi$=1. 

The  ${\rm H_2O}$ chemistry was discussed in detail by 
 \cite{vanDishoeck1986}, \cite{Hollenbach2009, Hollenbach2012}, and 
\cite{vanDishoeck2013}. As in the case of HF, the variations in the H$_2$O abundance reflect the competition between the formation and destruction processes that contribute over the parameter range explored here.  In brief, the destruction of
${\rm H_2O}$ is dominated by photodissociation and, at low $A_V$, has some contribution by
reaction with ${\rm C^+}$. 
The production of ${\rm H_2O}$ proceeds by a combination of surface chemistry
on grains and ion-neutral chemistry in the gas phase. When the molecular
 fraction is low the ion-neutral chemistry in the gas phase is initiated by
 cosmic-ray ionization of atomic H, while when the molecular fraction is high, the 
ion-neutral chemistry is initiated by cosmic-ray ionization of ${\rm H_2}$.
In both the atomic and molecular branches
the rate of production of ${\rm H_2O}$ increases with the ${\rm H_2}$ abundance
since both branches require reactions with ${\rm H_2}$ to proceed in the gas phase.
When both the ${\rm H_2}$  and electron abundances
are high, the production of water  can be reduced due to electron
recombinations with ${\rm H_3^+}$. When the ${\rm H_2}$ abundance
is low, but the electron abundance remains high, the water production can be reduced by electron recombination 
with ${\rm H^+}$ or by neutralization of ${\rm H^+}$ by ${\rm PAH^-}$ or by PAH. Finally, when the H$_2$ abundance is high, 
but the electron abundance is low, the gas-phase production of
H$_2$O proceeds efficiently through ion-molecule reactions.

To further explain the
physical regimes where the various processes dominate, we
start with a typical diffuse cloud visual extinction
($A_{V,{\rm tot}} =$ 0.5 mag), density ($n_{\rm H} = $100 ${\rm cm^{-3}}$), and cosmic-ray
ionization rate ($\zeta =$ 2$\times$10$^{-16}$ s$^{-1}$), and then discuss
the variation in the H$_2$O production, while varying these
parameters. For the $A_{V,{\rm tot}} \sim 0.5$ mag
and $n_{\rm H} \le 100$ ${\rm cm^{-3}}$ models, the production of
${\rm H_2O}$ is dominated by ion-neutral chemistry through the
cosmic-ray ionization of atomic hydrogen. To first order the
${\rm H_2O}$ abundance scales as $\zeta/n_{\rm H}$ and thus
increasing density  results in a lower $N({\rm H_2O})/N({\rm H_2})$ ratio, for a given cosmic-ray rate.

For models $n_{\rm H} \ge 300$ ${\rm cm^{-3}}$,
the molecular fraction rises and the atomic hydrogen abundance falls sufficiently
so that the ion-molecule chemistry proceeds mainly through cosmic-ray ionization
of ${\rm H_2}$. However, for these models, the electron abundance
remains high and the gas-phase production of ${\rm H_2O}$ is
reduced due to the recombination of ${\rm H_3^+}$,  an intermediary in this particular H$_2$O production route.
As a result,  ${\rm H_2O}$ is produced mainly by reactions on grains. In the limit of ${\rm H_2O}$ produced by grain chemistry alone and destroyed by photodissociation, the  $N({\rm H_2O})/N({\rm H_2})$ ratio is a 
function of  $n_{\rm H}/\chi$ (Hollenbach et al., 2009). 

For the $A_{V,{\rm tot}} \sim$0.5 mag, $n_{\rm H}=$300 ${\rm cm^{-3}}$, and $\zeta=$0.7$\times$10$^{-16}$ s$^{-1}$ model,
only $\sim 15$\% of the  water is predicted to be produced through gas-phase reactions.
For the lowest cosmic ray rate we
tested ($\zeta = 0.2\times 10^{-16}$ ${\rm s^{-1}}$),
the models with $n_{\rm H} \ge100$ ${\rm cm^{-3}}$ and $\chi = 1$
 are  dominated by grain surface chemistry
and increasing the density $n_{\rm H}$ produces more water and, hence, higher $N({\rm H_2O})/N({\rm H_2})$. 

At lower $A_{V,{\rm tot}}$ ($< 0.5$ mag),
curves for which grain surface chemistry dominates are seen to rise.
Since the grain surface chemistry reactions  do not depend on the ${\rm H_2}$ abundance,
the curves rise as the ${\rm H_2}$ abundance
drops. At higher $A_{V,{\rm tot}} \ge$ 2 mag, for all models, the electron abundance
drops sufficiently due to absorption of FUV photons, so that water
can be produced efficiently by ion-molecule chemistry.

We note that Dulieu et al. (2013) have presented experimental evidence
to suggest that OH and H$_2$O formed on bare silicate grain surfaces can be
chemically desorbed  upon formation. We consider the effects of chemical desorption by assuming that 30\% of the OH is immediately
desorbed and 70\% of the H$_2$O is immediately desorbed (see their Fig. 4). We find at most a $\sim$ 60\% increase in the H$_2$O column density over our model parameter space.
The effect is largest where surface chemistry dominates the production of H$_2$O and for the smallest cloud column densities. The increase drops to
$\sim$ 40\% for a total cloud column density of $A_{V,{\rm tot}}$ =2.0, and drops to
$\sim$ 10\% for a total cloud column density of $A_{V,{\rm tot}}$ = 4.0. We also note
that Minissale et al. (2013), and Minissale et al. (2014) have found that O-atom diffusion can be quite rapid on grain surfaces leading to the production of O$_2$ and O$_3$. This process is important deep in molecular clouds where the incidence of oxygen atoms exceeds that of H. These conditions are, however, not applicable to the
models presented here.

\subsection{H$_2$O Observations}

Since H$_2$ is not observable directly in the sub-mm, efforts are made to find surrogate tracers of H$_2$ in this wavelength domain in order to obtain abundance measurements relevant to test our knowledge of diffuse cloud chemistry. In Section 5, we demonstrated that HF can be used as a surrogate tracer of H$_2$, down to much lower molecular fractions than the more typical tracers such as CH. Since we do not have H$_2$  or CH column density measurements for all 47 features presented here, we use HF as a tracer of H$_2$ and discuss the H$_2$O abundance relative to HF -$N$(H$_2$O)/$N$(HF)- in the remainder of this paper. 

As described in Section 4, we detected absorption from H$_2$O toward all 12 sight lines included in our survey and we measured its column density in a total of 47 cloud components (see Table~\ref{tbl-3}). We compared our H$_2$O column density measurements against those reported by Flagey et al. (2013) for the 17 gas components in common to both studies; these are detected in the foreground to W49N, W51, G34.3$+$0.1, W33A and DR21(OH). We find that the two sets of measurements are consistent within 2 $\sigma$ for 13 components. For four components, the H$_2$O column densities we measured are either lower by 50\% (DR21(OH)) or higher by up to a factor of 3 (G34.3$+$0.1 and W49N) than those reported by Flagey et al. (2013). In all four cases, the background continua local to these gas components are somewhat affected by the presence of water emission local to the sources; the discrepancies we noted above are readily explained by the different methods we adopted to account for the background emission toward each source and give a measure of the modeling uncertainty in such cases. The H$_2$O column density measurements we report for these 4 velocity components (marked with an asterisk in Table~\ref{tbl-3}) should be used with caution.

Flagey  et al. (2013) reported no dependence of the H$_2$O abundance relative to H$_2$ on Galactocentric distance for the 6 sources they considered. For those sources located in the inner Galaxy in our extended survey, we computed a Galactocentric distance for each cloud we detected in HF and H$_2$O based on the source longitude and V$_{\rm LSR}$ to see whether the variations in the $N$(H$_2$O)/$N$(HF) we measure show a dependence on position in the Galaxy. We find no dependence of the $N$(H$_2$O)/$N$(HF) ratio with Galactic radius either, a conclusion consistent with that of the smaller Flagey et al. (2013) survey. Our results therefore add weight to the conclusion that the variations in the $N$(H$_2$O)/$N$(HF) ratio we measure can be considered representative throughout the Galactic disk. With a median value of $N$(H$_2$O)/$N$(HF) = 1.51 in diffuse clouds, H$_2$O also appears as an alternative tracer of H$_2$ (within a factor of 2.5) in the absence of HF or CH spectra, as already suggested in Flagey et al. (2013).

The comparison of our model predictions displayed in Fig. \ref{fig8} and Fig. \ref{fig9} shows that, in the diffuse molecular cloud regime ($f_{\rm H_2} \le 0.5$), the column density ratio of H$_2$O-to-HF varies from 0.5 $\le$ $N$(H$_2$O)/$N$(HF) $\le$ 3, all models considered.  Our models predict that the water abundance relative to HF varies by up to a factor 6, depending on the local gas physical conditions. The model comparisons further show that the variations in the $N({\rm H_2O})/N({\rm H_2})$ ratio are much larger (factor of 2 to 4)  than the variations in the
$N({\rm HF})/N({\rm H_2})$ ratio (at most factor of 2) both as a function of cloud column density for a given set of parameters and between models with different input parameters.  Our models further predict that the variations in the $N$(H$_2$O)/$N$(HF) are mostly driven by variations in the H$_2$O column density, rather than in the HF column densities for a given set of model parameters.

\begin{figure}
\epsscale{.9}
\plotone{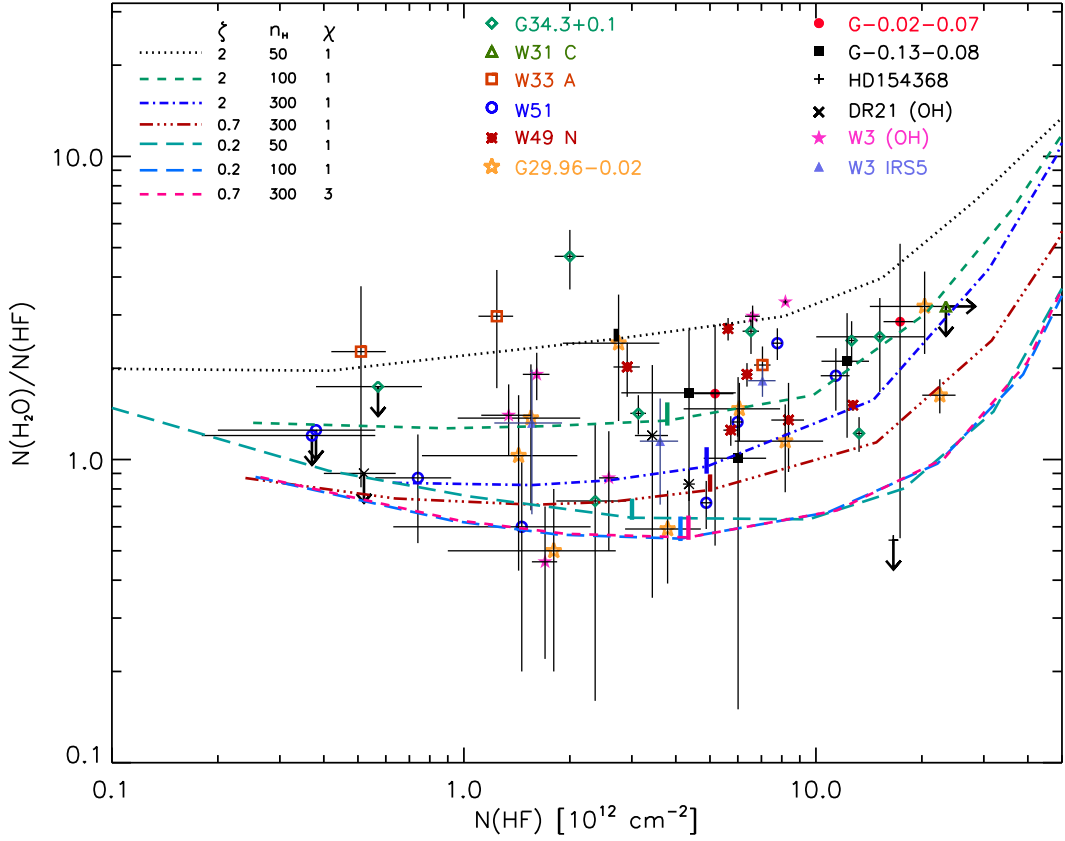}\caption{H$_2$O column density relative to that of HF versus HF column density throughout the Galactic disk probed by our {\it Herschel}  survey. Each sight line is represented by a particular symbol and color which are listed at the top of this figure.  Overplotted on our measurements are the predictions obtained from a series of 2-sided PDR model runs. The parameter space explored in our runs is listed in the upper left of the panel with $\zeta$ in units of 10$^{-16}$ s$^{-1}$ and $n_{\rm H}$ in units of cm$^{-3}$. 1 $\sigma$ uncertainties are over plotted for each measurement. 3-$\sigma$ uncertainties are displayed for all upper limits. The colored vertical bars over plotted on each model run correspond to a depth in the cloud of $A_{V,\rm tot} \sim 0.5$ mag. The upper limit for HD154368 was derived using H$_2$O results from Spaans et al. (1998) and HF and H$_2$ results from Indriolo et al. (2013). \label{fig10}}
\end{figure}

In Figure~\ref{fig10} we display the $N$(H$_2$O)/$N$(HF) ratio {\it versus} $N$(HF) for all 47 foreground clouds with optically thin or moderately thick absorptions. The measurements toward each sight line are represented by a particular symbol and color that are listed at the top of the panel.  We over-plot our model predictions for the $N$(H$_2$O)/$N$(HF) ratio onto our measurements using the same symbol and color coding as in Figs~\ref{fig8} \&\ref{fig9}. The range of model parameters we use is listed at the top left of the panel. We use a log-log scale for both axes for clarity purposes. 1$\sigma$ uncertainties are shown for all measurements and 3$\sigma$ uncertainties are plotted as downward arrows for upper limits in the water abundance. This figure shows that the $N$(H$_2$O)/$N$(HF) ratio varies by about a factor 5 overall; our observational results are therefore fully consistent with the model predictions presented here and known to best bracket the physical conditions in diffuse clouds.

Our sight line analyses (Section 2) indicated that HF and H$_2$O are co-located in the gas components we detect throughout the Galactic disk, meaning that both molecules are subject to the same physical conditions within each gas clump along the sight lines. As a result, the variations we measure are not caused by differences in the spatial distributions of HF or H$_2$O along a given sight line (geometric effects) but are caused by true variations in the molecular content for a given gas component.

In particular, Fig. \ref{fig10} displays a subtle break in the water abundance relative to HF once $N$(HF)$=$5$\times$10$^{12}$ cm$^{-2}$. For $N$(HF) $\leq$ 5$\times$10$^{12}$ cm$^{-2}$, our measurements indicate that the $N$(H$_2$O)/$N$(HF) ratio varies by about a factor of 6 to 9 from $\sim$ 0.5 to 4.6. The lowest water abundances relative to HF are best reproduced by models with low cosmic ray ionization rates ($\zeta \leq 0.7\times$10$^{-16}$ s$^{-1}$) over the density range known to be representative of the diffuse ISM while the highest water abundances relative to HF are best represented by $\zeta \sim 2\times$10$^{-16}$ s$^{-1}$ and $n_{\rm H} \leq$ 100 cm$^{-3}$. The prevalence of mixtures of physical conditions within the galactic disk diffuse ISM and along any given  sightline has long been recognized and is clearly evidenced here again, as multiple model runs can accommodate our measurements for this particular HF column density range. 

For $N$(HF) $>$ 5$\times$10$^{12}$ cm$^{-2}$, however, our measurements are not consistent with models using $\zeta \leq 0.7\times$10$^{-16}$  s$^{-1}$. As a matter of fact, over 80\% of our measurements in this specific HF column density range show water abundances relative to HF consistent with model predictions for single clouds with $\zeta = 2\times$10$^{-16}$ s$^{-1}$, $\chi=$1 and n$_{\rm H} \leq$100 cm$^{-3}$. The variations in the measured $N$(H$_2$O)/$N$(HF) in this HF regime amount to at most a factor of 3, thus, pointing toward a population of low gas density clouds with properties intrinsically very similar.  Considering that the HF abundance relative to H$_2$ currently measured varies by at most a factor of $\sim$2 (see Section 5), we conclude that the variations we measured in the $N$(H$_2$O)/$N$(HF) ratio are mostly driven by variations in the H$_2$O/H$_2$ ratio within the gas clouds we probe.

\section{Discussion}

In the diffuse ISM, the production of H$_2$O within our models proceeds via both gas-phase reactions and grain-surface chemistry. Since the contribution of each mechanism depends on the cloud physical conditions, our measurements give us a unique opportunity to test the chemical pathways predicted to lead to H$_2$O formation in this regime.

\begin{figure}
\epsscale{.9}
\plotone{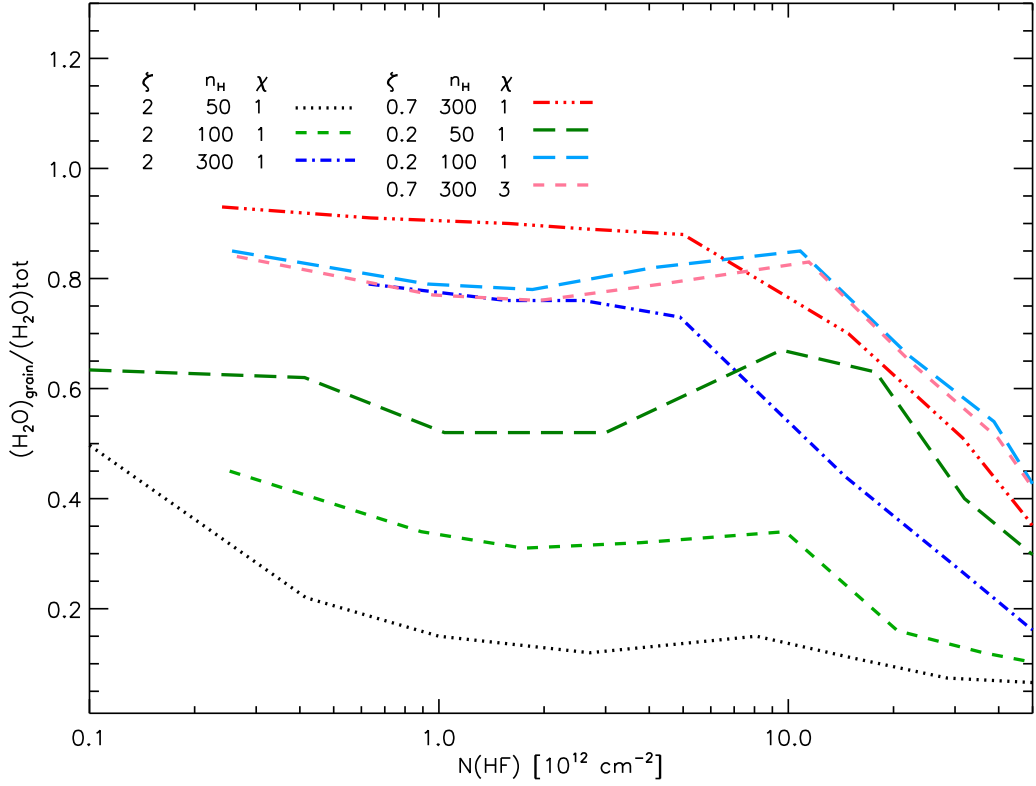}\caption{Fractional rate of H$_2$O production on grain surfaces compared to the total production rate of H$_2$O {\it versus}  HF column density in the column density range probed by the survey. The parameter space explored in our runs is listed in the upper section of the panel with $\zeta$ in units of 10$^{-16}$ s$^{-1}$ and $n_{\rm H}$ in units of cm$^{-3}$. The UV field is given in units of the Draine (1978) field and is fixed to $\chi=$1 or $\chi=$3. \label{fig12}}
\end{figure}

In Figure \ref{fig12},  we show the fractional rate of H$_2$O production on grain
surfaces compared to the total production rate of H$_2$O as a function
of $N$(HF). The rates are evaluated at cloud center where the H$_2$O abundance
is generally highest and dominates the contribution to the column density.
These curves can be directly compared with those in Figure \ref{fig10}.

As already mentioned in Section 6, at a column of $N$(HF) $\approx$ 5$\times$10$^{12}$ cm$^{-2}$
and for $n_{\rm H} \le$ 100 cm$^{-3}$, the production of H$_2$O is dominated
by ion-neutral chemistry (see Fig.\ref{fig12}). The H$_2$O abundance scales
as $\zeta$/$n_{\rm H}$ and thus a larger density results in a lower
$N$(H$_2$O)/$N$(HF) ratio for a given cosmic-ray rate (Fig. \ref{fig10}).
At higher densities $n_{\rm H} >$ 300 cm$^{-3}$, grain-surface reactions dominate
the production of water. Figure \ref{fig12} shows that for $n_{\rm H} =$ 300 cm$^{-3}$
and $\zeta =$ 0.7$\times$10$^{-16}$ s$^{-1}$ only about 15\% of the water is produced
through gas-phase reactions.

For $N({\rm HF})$ $< 5\times 10^{12}$ ${\rm cm^{-2}}$ ($A_{V,\rm tot}$ $< 0.5$ mag) ,
the curves for which grain surface chemistry dominates are seen to rise in Fig. 10.
Since the grain chemistry reactions  do not depend on the ${\rm H_2}$ abundance
while the HF production does, the curves rise as the ${\rm H_2}$ and thus HF abundance
drops. At higher $A_{V,\rm tot}$, the electron abundance drops sufficiently due to absorption of 
FUV photons, so that water can then be produced
efficiently through the cosmic-ray ionization of ${\rm H_2}$. The curves thus rise
due to an increasing abundance of ${\rm H_2O}$.

Note that in the limit in which ${\rm H_2O}$ is produced by grain surface chemistry alone and destroyed
by photodissociation, while HF is both formed and destroyed by gas-phase
reactions, the ratio $N({\rm H_2O})/N({\rm HF})$ is a function of  $n_{\rm H}/\chi$ 
(see Figs. \ref{fig10} \& \ref{fig12}). For the lowest cosmic ray rate we
considered here ($\zeta = 0.2\times 10^{-16}$ ${\rm s^{-1}}$), 
the curve for $n_{\rm H}=100$ ${\rm cm^{-3}}$ and $\chi = 1$ lies close to the curve for $\zeta = 0.7\times 10^{-16}$ ${\rm s^{-1}}$, $n_{\rm H} = 300$ ${\rm cm^{-3}}$ and $\chi = 3$,  as expected for  ${\rm H_2O}$ production by grain chemistry (see Fig.~\ref{fig12}).

Our measured water abundances relative to HF reported in Table~\ref{tbl-3} are all bracketed by model predictions based on a grid of gas physical conditions known to be representative of the diffuse ISM.  At low visual extinctions and for densities greater than $n_{\rm H}=$100 cm$^{-3}$, our models predict that grain-surface chemistry is necessary to account for some of our observations. For $A_{V,\rm tot}$ $\geq$ 0.5 mag, our models indicate that contributions from both gas-phase and grain-surface chemistry are required to account for our measurements as well. The contribution of each production mechanism depends intimately on the electron density distribution within the cloud, as recombination with H$^+$ and H$_3^+$ inhibit the gas-phase production routes.

Within our uncertainties, most measurements are compatible with more than one model. 
These degeneracies are mainly due to two factors. Firstly, some of our model predictions are degenerate and measurements of HF and H$_2$O alone do not allow us to lift the model degeneracies; additional observational probes such as HCO$^+$ (e.g., Godard et al. 2014) are necessary to do so. Secondly, our measurements are not obtained toward true single clouds contrary to the model assumptions, as discussed earlier. 
To compare the water abundance relative to HF {\it versus} HF column density on equal footing one would have to determine the HF column density in each gas clump along a given sight line, a task which is observationally very challenging. One rare exception might be the sight line toward HD154368. As mentioned earlier, Spaans et al. (1998) and Rachford et al. (2002) determined that true translucent conditions are present in the gas intercepting the sight line to this star. The decomposition performed by Rachford et al. (2002) suggests that $\sim$73\% of the H$_2$ gas observed with {\it FUSE} data belongs to the translucent phase.  As a result, it is reasonable to assume that $\sim$73\% of the total HF column density we measure is also associated with this translucent phase. Correcting the total measured HF column density by that fraction has the effect of moving the upper limit on the $N$(H$_2$O)/$N$(HF) ratio to the left (or $N$(HF)$\sim$ 1.2 $\times$10$^{13}$ cm$^{-2}$) in the bottom panel of Fig.~\ref{fig10}.  One can see that the overall impact of such a  sight line ``decomposition'' on our measurements would primarily be to shift the distribution horizontally and to the left - or to lower $N$(HF). 

Despite these degeneracies, it is important to note that the lower envelope on the $N$(H$_2$O)/$N$(HF) ratios we measure is too high to understand without grain-surface chemistry being important. The overall range in $N$(H$_2$O)/$N$(HF) ratios that we observe in itself demonstrates the importance of grain surface chemistry in the production of water in this diffuse ISM.  
It is also remarkable to see that the adopted physical conditions and chemical network that best bracket our observations remain largely unaltered, when taking the latter distribution shift into account. Our measurements therefore indicate that the range in densities, FUV and cosmic-ray ionization rates required to model our observations are not only pervasive throughout the diffuse gas composing the Galactic disk but they are also quite tightly constrained. Finally, our results also validate our understanding of the various chemical pathways thought to lead to the gas-phase abundance of HF and H$_2$O in the diffuse ISM. While most chemical pathways involve gas phase reactions in diffuse clouds, our measurements do confirm that grain surface chemistry also plays a significant role in the production of gas-phase water in such low density environments.

We note that the effects of turbulent dissipation regions (TDR) on the chemistry in diffuse clouds were recently extensively tested and were found to successfully reproduce the wealth of CH$^+$ and SH$^+$ absorption measurements obtained with {\it Herschel} toward sight lines probing the diffuse ISM (Godard et al. 2014 and references therein). The production of gas-phase water was also included in those models. Godard et al. (2014) found that for gas densities in the range we probe here ($n_{\rm H}$= 50-300 cm$^{-3}$) and for $A_{V,\rm tot}$$\sim$0.5 mag and a primary cosmic ray rate per H of 0.4$\times$10$^{-16}$ s$^{-1}$ cm$^{-3}$,  TDR could contribute up to 50\% of the total (TDR plus ion-molecule) gas-phase production of water. The Godard et al. (2014) models do not include the effect of grain-surface chemistry, while our models do not include the effects of TDR on the water production, hence a direct comparison of these model predictions is not feasible. However, all differences considered and for this particular set of parameters, our models predict that grain surface chemistry produces up to 85\% of the observed gas-phase water abundance, with ion-molecule reactions producing 15\% of the observed gas-phase water abundance. Consequently, ion-molecule reactions combined with TDR effects alone (without any contribution from grain surface chemistry) would fall short ($\sim$35\% of that required) in terms of reproducing the observed water abundance measured in this survey.

\section{Conclusions}

We have presented the most comprehensive survey of HF and H$_2$O observations obtained in the diffuse interstellar medium of the Galactic disk using the {\it Herschel}/HIFI instrument. The column density of both molecules is measured toward 47 discrete gas components detected in absorption and probing the Galactic disk volume. We demonstrate that both molecules are ubiquitous in the diffuse gas of our Galaxy.  We find that the HF and H$_2$O velocity distributions trace each other almost perfectly, in the disk, establishing that HF and H$_2$O essentially probe the same gas-phase volume. 

We compare our observations to state-of-the-art diffuse cloud models that were modified to include grain surface chemistry, as well as updated reaction rates for the fluorine chemistry. Our measurements corroborate theoretical predictions that HF is a very sensitive tracer of H$_2$ down to molecular fractions of a few percent, a regime where more common tracers such as CH are below detection level. 

We have used HF as a surrogate tracer of H$_2$  to study the variation of the H$_2$O column density -relative to HF- within the Galactic disk diffuse gas. We find that the $N$(H$_2$O)/$N$(HF) ratio shows a narrow distribution with a median water abundance relative to HF of 1.51.  Our results therefore add weight to the previous suggestion that H$_2$O can also be used as tracer of H$_2$- within a factor of 2.5-  in the diffuse ISM,  in the absence of HF or CH observations.

We have demonstrated that the overall variations of a factor of 2.5 around the median in the $N$(H$_2$O)/$N$(HF) ratio are driven by true variations in the H$_2$O column density within the disk. Our measurements, therefore, provide us with a unique opportunity to test the chemical pathways predicted to play a role in the water production in the diffuse ISM. We find that the range in water abundances relative to HF that we measure could only be explained if significant grain surface chemistry production occurred in addition to gas-phase ion-molecule production. While most chemical pathways involve gas phase reactions alone in the diffuse ISM, our survey confirms that grain surface chemistry can play a significant role in the production of some molecular species, such as gas phase H$_2$O,  in this low density environment.  



\acknowledgments
We thank the referee for very useful comments which improved the manuscript significantly. We thank David Hollenbach for useful comments on grain surface chemistry. P.S. acknowledges support for this work from NASA through award NNN12AA01C issued by JPL/Caltech. M.G.W. was supported in part by NSF grant AST-1411827. MG thanks CNES and PCMI/INSU for support for the analysis of the
Herschel data. This work was carried out in part at the Jet Propulsion Laboratory, which is operated for NASA by the California Institute of Technology. HIFI has been designed and built by a consortium of institutes and university departments from across Europe, Canada and the United States under the leadership of SRON Netherlands Institute for Space Research, Groningen, The Netherlands and with major contributions from Germany, France and the US. Consortium members are: Canada: CSA, U.Waterloo; France: CESR, LAB, LERMA, IRAM; Germany: KOSMA, MPIfR, MPS; Ireland, NUI Maynooth; Italy: ASI, IFSI-INAF, Osservatorio Astrofisico di Arcetri-INAF; Netherlands: SRON, TUD; Poland: CAMK, CBK; Spain: Observatorio Astronomico Nacional (IGN), Centro de Astrobiolog'a (CSIC-INTA). Sweden: Chalmers University of Technology - MC2, RSS \& GARD; Onsala Space Observatory; Swedish National Space Board, Stockholm University - Stockholm Observatory; Switzerland: ETH Zurich, FHNW; USA: Caltech, JPL, NHSC.



{\it Facilities:} \facility{Herschel}.

\clearpage

\clearpage

\begin{table}
\caption{Target List. \label{tbl-1}}
\vspace{0.5cm}
\begin{tabular}{lcccccc}
\hline\hline
Target & RA & Dec  & $l$ & $b$ & $D$ & Ref.  \\
	& ($^h$) ($^m$) ($^s$)	& ($^\circ$) ($^\prime$)($^{\prime\prime}$)	&	($^\circ$)	&	($^\circ$)	&	(kpc) &	\\  \hline
W3 IRS5	&	02 \,25 \,40.6	&	$+$62 \,05 \,51.0	&	133.715	&	$+$1.215	&	1.83 & 1	\\
W3(OH)	&	02 \,27 \,03.8	&	$+$61 \,52 \,25.0	&	133.946	&	$+$1.06	&	2.04 &  2	\\
G$-$0.02$-$0.07	&	17 \,45 \,50.2	&	$-$28 \,59 \,53.0	&	359.97	&	$-$0.07	&	8.34	& 3\\
G$-$0.13$-$0.08	&	17 \,45 \,37.4	&	$-$29 \,05 \,40.0	&	359.30	&	$-$0.432	&	8.34	& 3\\
G34.3+0.1	&	18 \,53 \,18.7	&	$+$01 \,14 \,58.0	&	34.26	&	$+$0.15	&	3.80 & 4	\\
W28A	&	18 \,00 \,30.4	&	$-$24 \,04 \,00.0	&	5.9	&	$-$0.39	&	1.28	& 5\\
W31C	&	18 \,10 \,28.7	&	 $-$19 \ 55 \,50.0 	&	10.62	&	$-$0.38	&	4.95	& 6\\
W33A 	&	18 \,14 \,39.4 	&	$-$17 \,52 \,00.0	&	12.91	&	 $-$0.26	&	2.40	& 7\\
G29.96$-$0.02	&	18 \,46 \,03.9	&	$-$02 \,39 \,21.9	&	29.96	&	$-$0.02	&	5.26	& 8 \\
W49N 	&	 19 \,10 \,13.2 	&	$+$09 \,06 \,12.0	&	43.17	&	$+$0.01	&	11.1	& 9\\
W51 	&	 19 \,23 \,43.9 	&	$+$14 \,30 \,31.0 	&	49.49	&	 $-$0.39	&	5.41	& 10\\
DR21(OH)  	&	20 \,39 \,01.0 	&	$+$42 \,22 \,48.0	&	81.72	&	$+$0.57	&	1.50	& 11\\
\hline\hline									
\end{tabular}\\

References: 1- Imai et al. (2000); 2- Hachisuka et al. (2006) ; 3- Reid et al. (2014); 
4- Fish et al. (2003); 5- Motogi et al. (2011); 6- Sanna et al. (2014); 7- Immer et al. (2013); 
8- Zhang et al. (2014); 9- Zhang et al. (2013) ; 10- Sato et al. (2010); 11- Rygl et al. (2012) 	
\end{table}



\clearpage

\begin{deluxetable}{lccccll}
\tabletypesize{\scriptsize}
\rotate
\tablecaption{Observation Summary.\label{tbl-2}}
\tablewidth{0pt}
\tablehead{
\colhead{Target} & \colhead{t$_{exp}$(HF)} & \colhead{T$_A$(cont)[HF]$^a$} & \colhead{t$_{exp}$(H$_2$O)} & \colhead{T$_A$(cont)[H$_2$O]$^a$} &
\colhead{Observation Dates} & \colhead{HF \& H$_2$O Obs IDs} \\
& (s) & (K) & (s) & (K) & & starting  with 1342:   
}
\startdata
\hline													\\
W3 IRS5	&	492	&	3.56 $\pm$ 0.10	&	2965	&	3.11 $\pm$ 0.02	&	2012-07-20 \& 2010-07-18  &	248382-83-84 \& 201591	\\
W3(OH)	&	9275	&	4.46 $\pm$ 0.03	&	3540	&	4.14 $\pm$ 0.03	&	2013-03-27 \& 28	&	268474-75-76 \& 268605-06-07 	\\
G$-$0.02$-$0.07	&	3693	&	0.93 $\pm$ 0.04	&	2139	&	0.92 $\pm$ 0.04	&	2010-10-06	&	205885-86-87 \& 205882-83-84	\\
G$-$0.13$-$0.08	&	746	&	1.43 $\pm$ 0.11	&	428	&	1.39 $\pm$ 0.11	&	2011-09-14	&	228613-14  \& 228615-16 	\\
G34.3+0.1	&	522	&	9.20 $\pm$ 0.12	&	246	&	7.57 $\pm$ 0.10	&	2010-04-18	&	195074-75-76 \& 195070-71-72	\\
W28A	&	562	&	6.03 $\pm$ 0.10	&	426	&	4.71 $\pm$ 0.06	&	2011-03-11	&	215860-61-62 \& 215863-64-65	\\
W31C	&	534	&	7.74 $\pm$ 0.11	&	306	&	5.99 $\pm$ 0.09	&	2010-03-05	&	191690-91-92 \& 191687-88-89	\\
W33A 	&	3775	&	2.36 $\pm$ 0.04	&	2694	&	1.91 $\pm$ 0.03	&	2011-03-11	&	215875-76-77 \& 215872-73-74	\\
G29.96$-$0.02	&	9162	&	3.20 $\pm$ 0.03	&	2824	&	2.69 $\pm$ 0.01	&	2013-03-27 \& 2011-09-29	&	268483-84-85 \& 229875-76	\\
W49N 	&	522	&	11.48 $\pm$ 0.18	&	246	&	9.06 $\pm$ 0.16	&	2010-03-22	&	192595-96-97 \& 192592-93-94	\\
W51 	&	9259	&	11.12 $\pm$ 0.04	&	3582	&	10.51 $\pm$ 0.04	&	2013-03-27 \& 28	&	268478-79-80 \& 268611-12-13	\\
DR21(OH)  	&	1014	&	5.36 $\pm$ 0.06	&	1086	&	4.56 $ \pm$ 0.05	&	2010-05-12	&	196502-03-04 \&196506-07-08   	\\
\hline													
\enddata
\tablenotetext{a}{T$_A$(cont) is the double sideband continuum antenna temperature.}
\end{deluxetable}

\clearpage

\begin{deluxetable}{lcrrrr}
\tablewidth{415pt}
\tablecaption{Column Density Measurements. \label{tbl-3}}
\tablehead{
\colhead{$V_{\rm LSR}$ Range}       &  \colhead{$N$(HF)}          & \colhead{$N$(p-H$_2$O)}  &
\colhead{$N$(H$_2$O)$_{\rm tot}$} & \colhead{$\frac{N({\rm H_2O})_{\rm tot}}{N({\rm HF})}$} \\
\scriptsize{(km s$^{-1}$)}    & \scriptsize{(10$^{12}$ cm$^{-2}$)} & \scriptsize{(10$^{12}$  cm$^{-2}$)} & \scriptsize{(10$^{12}$ cm$^{-2}$)} & } 
\startdata
\sidehead{\bf W49N} 
$+$30,$+$34*	& 	12.74$\pm$	0.22	&	4.82	$\pm$	0.16	&	19.27	$\pm$	0.66	&	1.51	$\pm$	0.06	\\
$+$43,$+$48	& 	2.91	$\pm$	0.25	&	1.47	$\pm$	0.27	&	5.87	$\pm$	1.07	&	2.02	$\pm$	0.41	\\
$+$49,$+$53*	& 	5.73	$\pm$	0.27	&	1.80	$\pm$	0.18	&	7.19	$\pm$	0.73	&	1.25	$\pm$	0.14	\\
$+$53,$+$55	& 	6.37	$\pm$	0.24	&	3.04	$\pm$	0.25	&	12.18	$\pm$	1.01	&	1.91	$\pm$	0.17	\\
$+$55,$+$58	& 	5.63	$\pm$	0.22	&	3.80	$\pm$	0.28	&	15.22	$\pm$	1.11	&	2.70	$\pm$	0.23	\\
$+$64,$+$71	& 	8.36	$\pm$	0.89	&	2.82	$\pm$	0.88	&	11.27	$\pm$	3.51	&	1.35	$\pm$	0.44	\\ \tableline 
\sidehead{\bf W51} 
$+$4,$+$6	& 	4.88	$\pm$	0.17	&	0.88	$\pm$	0.16	&	3.53	$\pm$	0.62	&	0.72	$\pm$	0.13	\\
$+$6,$+$10	& 	11.39$\pm$	1.07	&	5.37	$\pm$	0.11	&	21.49	$\pm$	4.54	&	1.89	$\pm$	0.44	\\
$+$11,$+$15	& 	1.46	$\pm$	0.83	&	0.22	$\pm$	0.08	&	0.88 $\pm$	0.32	&	0.60	$\pm$	0.40	\\
$+$19,$+$21	& 	0.38	$\pm$	0.18	&	$<$ 0.12		&	$<$ 0.48			&	$<$ 1.25 	\\
$+$21,$+$24	& 	0.37	$\pm$	0.19	&	$<$ 0.11		&	$<$ 0.44			&	$<$ 1.20 	\\
$+$24,$+$27	& 	0.74	$\pm$	0.18	&	0.16	$\pm$	0.05	&	0.64	$\pm$	0.20	&	0.87	$\pm$	0.34	\\
$+$40,$+$46	& 	7.77	$\pm$	0.25	&	4.71	$\pm$	0.54	&	18.82	$\pm$	2.15	&	2.42	$\pm$	0.29	\\
$+$47,$+$50	& 	5.99	$\pm$	0.20	&	1.99	$\pm$	0.20	&	7.95	$\pm$	0.79	&	1.33	$\pm$	0.14	\\ \tableline
\sidehead{\bf G29.96$-$0.02} 
$+$3,$+$5		& 	1.55	$\pm$	0.59	&	0.53	$\pm$	0.17	&	2.12	$\pm$	0.70	&	1.37	$\pm$	0.69	\\
$+$7,$+$11		&   20.36	$\pm$	6.13	&	16.31	$\pm$	0.47	&	65.22	$\pm$	1.86	&	3.20	$\pm$	0.97	\\
$+$11,$+$15	& 	6.06	$\pm$	1.85	&	2.22	$\pm$	0.23	&	8.88	$\pm$	0.92	&	1.47	$\pm$	0.32	\\
$+$16,$+$18	& 	1.43	$\pm$	0.67	&	0.37	$\pm$	0.14	&	1.48	$\pm$	0.56	&	1.03	$\pm$	0.60	\\
\sidehead{\bf G29.96$-$0.02} 
$+$51,$+$56	& 	3.79	$\pm$	0.92	&	0.56	$\pm$	0.13	&	2.22	$\pm$	0.52	&	0.59	$\pm$	0.20	\\
$+$57,$+$62	&   8.18	$\pm$	2.30	&	2.35	$\pm$	0.35	&	9.41	$\pm$	1.40	&	1.15	$\pm$	0.37	\\
$+$65,$+$73	& 	22.49$\pm$	2.44	&	9.17	$\pm$	0.59	&	36.70	$\pm$	2.35	&	1.63	$\pm$	0.21	\\
$+$74,$+$77	& 	1.80	$\pm$	0.90	&	0.22	$\pm$	0.08	&	0.90	$\pm$	0.30	&	0.50	$\pm$	0.30	\\
$+$89,$+$93	& 	2.75	$\pm$	0.84	&	1.67	$\pm$	0.55	&	6.67	$\pm$	2.18	&	2.42	$\pm$	1.08	\\  \tableline			
\sidehead{\bf G34.3$+$0.1} 
$+$8,$+$10 		& 	0.57	$\pm$	0.19	&	$<$ 0.25		&	$<$1.00		&	$<$ 1.74		\\
$+$10,$+$13	& 	15.19	$\pm$	5.17	&	9.66	$\pm$	0.46	&	38.62	$\pm$	1.85	&	2.54	$\pm$	0.87	\\
$+$13,$+$16	& 	3.13	$\pm$	0.16	&	1.11	$\pm$	0.16	&	4.43	$\pm$	0.62	&	1.42	$\pm$	0.21	\\
$+$17,$+$26	& 	2.36	$\pm$	0.53	&	0.43	$\pm$	0.32	&	1.71	$\pm$	1.28	&	0.73	$\pm$	0.57	\\
$+$27,$+$30*	& 	12.63	$\pm$	0.39	&	7.79	$\pm$	1.22	&	31.17	$\pm$	4.87	&	2.47	$\pm$	0.39	\\
$+$40,$+$44	& 	2.00	$\pm$	0.19	&	2.34	$\pm$	0.47	&	9.36	$\pm$	1.89	&	4.68	$\pm$	1.04	\\
$+$44,$+$47	& 	6.54	$\pm$	0.35	&	4.33	$\pm$	0.64	&	17.32	$\pm$	2.56	&	2.65	$\pm$	0.42	\\
$+$48,$+$50	& 	13.26	$\pm$	0.33	&	4.05	$\pm$	0.52	&	16.20	$\pm$	2.08	&	1.22	$\pm$	0.16	\\ \tableline				
\sidehead{\bf W33A} 
$+$20,$+$23	& 	0.51	$\pm$	0.09	&	0.29	$\pm$	0.18	&	1.15	$\pm$	0.71	&	2.27	$\pm$	1.46	\\
$+$24,$+$27	& 	7.05	$\pm$	0.38	&	3.61	$\pm$	0.51	&	14.45	$\pm$	2.05	&	2.05	$\pm$	0.31	\\
$+$43,$+$46	& 	1.24	$\pm$	0.14	&	0.92	$\pm$	0.37	&	3.69	$\pm$	1.49	&	2.97	$\pm$	1.25	\\ 
\sidehead{\bf W31C}  
$+$41,$+$47	& 	$>$23.4			&	18.58	$\pm$	0.35	&	74.32	$\pm$	1.39	&	$<$3.2	\\	
\sidehead{\bf DR21(OH)} 	
$-$11,$-$9& 	0.52	$\pm$	0.12	&	$<$0.12			&	$<$ 0.47		&	$<$ 0.90		\\
$+$11,$+$13*	& 	4.36	$\pm$	0.17	&	0.90	$\pm$	0.11	&	3.59	$\pm$	0.43	&	0.83	$\pm$	0.10	\\
$+$13,$+$16	& 	3.43	$\pm$	0.37	&	1.03	$\pm$	0.72	&	4.11	$\pm$	2.89	&	1.20	$\pm$	0.85	\\  \tableline		
\sidehead{\bf W3(OH)} 
$-$22,$-$18	& 	1.70	$\pm$	0.14	&	0.19	$\pm$	0.10	&	0.76 $\pm$ 0.40	& 0.45 $\pm$ 0.24	\\
$-$13,$-$8	& 	6.61	$\pm$	0.32	&	4.90	$\pm$	0.29	&	19.61	$\pm$	1.14	&	2.97	$\pm$	0.25	\\
$-$6,$-$4	& 	1.34	$\pm$	0.22	&	0.47	$\pm$	0.10	&	1.89	$\pm$	0.38	&	1.40	$\pm$	0.37	\\
$-$4,$-$2	& 	1.61	$\pm$	0.14	&	0.77	$\pm$	0.12	&	3.08	$\pm$	0.47	&	1.91	$\pm$	0.34	\\
$-$2,$+$0	 	& 	2.58	$\pm$	0.16	&	0.56	$\pm$	0.23	&	2.25	$\pm$	0.93	&	0.87	$\pm$	0.37	\\
$+$0,$+$2		& 	8.19	$\pm$	0.11	&	6.78	$\pm$	0.08	&	27.13	$\pm$	0.31	&	3.31	$\pm$	0.06	\\ \tableline
\sidehead{\bf W3 IRS5}  
$-$22,$-$17	& 	7.05	$\pm$	0.63	&	3.20	$\pm$	0.24	&	12.82	$\pm$	0.94	&	1.82	$\pm$	0.21	\\
$-$6,$-$3	& 	1.56	$\pm$	0.34	&	0.51	$\pm$	0.23	&	2.06	$\pm$	0.92	&	1.32	$\pm$	0.66	\\
$-$2,$+$2		& 	3.61	$\pm$	0.45	&	1.04	$\pm$	0.38	&	4.17	$\pm$	1.50	&	1.15	$\pm$	0.44	\\ \tableline
\sidehead{\bf G$-$0.02$-$0.07}  
$-$46,$-$37	& 	17.34	$\pm$	1.78	&	12.34	$\pm$	9.89	&	49.37	$\pm$	39.57	&	2.85	$\pm$	2.30	\\
$+$13,$+$18		& 	5.17	$\pm$	0.66	&	2.14	$\pm$	1.44	&	8.55	$\pm$	5.74	&	1.65	$\pm$	1.13	\\  \tableline
\sidehead{\bf G$-$0.13$-$0.08} 
$-$45,$-$41	& 	6.01	$\pm$	1.21	&	1.52	$\pm$	1.15	&	6.09	$\pm$	4.58	&	1.01	$\pm$	0.86	\\
$+$20,$+$25		& 	12.26	$\pm$	1.89	&	6.47	$\pm$	2.68	&	25.88	$\pm$	10.74	&	2.11	$\pm$	0.93	\\
$+$26,$+$30		& 	4.36	$\pm$	1.56	&	1.82	$\pm$	0.90	&	7.26	$\pm$	3.62	&	1.66	$\pm$	1.02	\\ \tableline
\enddata
\tablenotetext{a}{$N$(H$_2$O)$_{\rm tot}$ was derived from our $N$(p-H$_2$O) measurements assuming an ortho-para ratio of 3 for water (Flagey et al. 2013).}
\tablenotetext{*}{Our $N$(p-H$_2$O) value is higher by up to a factor 3 than that reported by Flagey et al. (2013) toward W49N and G34.3$+$0.1; our $N$(p-H$_2$O) value is lower by $\sim$ 50\%  than that reported by Flagey et al. (2013) toward DR21(OH). Details can be found in Section 6.2.}

\end{deluxetable}




\end{document}